\documentclass[aps,prd,onecolumn, tightenlines,superscriptaddress,
  amsmath,amssymb,eqsecnum,nofootinbib
]{revtex4-2}

\usepackage[utf8]{inputenc}
\usepackage{lmodern}
\usepackage{graphicx}
\usepackage{bm}
\usepackage[hidelinks]{hyperref}

\hypersetup{
  pdftitle={Vacuum stability in Geometric Trinity of Gravity},
  pdfauthor={Vincenzo Branchina, Salvatore Capozziello, Filippo Contino, Carmen Ferrara}
}

\newcommand{\be}{\begin{equation}}
\newcommand{\ee}{\end{equation}}
\newcommand{\bea}{\begin{eqnarray}}
\newcommand{\eea}{\end{eqnarray}}
\newcommand{\gr}[1]{\overset{\circ}{#1}{}}
\newcommand{\qm}[1]{``#1''}
\newcommand{\vv}{``}

\def\nn{\nonumber}
\newcommand{\dd}{\mathrm{d}}

\begin{document}

\title{Vacuum stability in Geometric Trinity of Gravity}

\author{Vincenzo Branchina}
\email{branchina@ct.infn.it}
\affiliation{Dipartimento di Fisica e Astronomia ``E. Majorana'', Universit\`a di Catania, Via S. Sofia 64, I-95123 Catania, Italy}
\affiliation{INFN Sezione di Catania, Via S. Sofia 64, I-95123 Catania, Italy}

\author{Salvatore Capozziello}
\email{capozziello@na.infn.it}
\affiliation{Scuola Superiore Meridionale, Via Mezzocannone 4, I-80134 Napoli, Italy}
\affiliation{Dipartimento di Fisica ``E.\ Pancini'', Universit\`a di Napoli ``Federico II'', Via Cinthia 9, I-80126 Napoli, Italy}
\affiliation{INFN Sezione di Napoli, Complesso Universitario di Monte Sant'Angelo, Edificio 6, Via Cinthia, I-80126, Napoli, Italy}

\author{Filippo Contino}
\email{f.contino@ssmeridionale.it}
\affiliation{Scuola Superiore Meridionale, Via Mezzocannone 4, I-80134 Napoli, Italy}
\affiliation{INFN Sezione di Napoli, Complesso Universitario di Monte Sant'Angelo, Edificio 6, Via Cinthia, I-80126, Napoli, Italy}

\author{Carmen Ferrara}
\email{carmen.ferrara-ssm@unina.it}
\affiliation{Scuola Superiore Meridionale, Via Mezzocannone 4, I-80134 Napoli, Italy}
\affiliation{INFN Sezione di Napoli, Complesso Universitario di Monte Sant'Angelo, Edificio 6, Via Cinthia, I-80126, Napoli, Italy}

\begin{abstract}
We investigate whether the classical equivalence of general relativity, its teleparallel equivalent, and its symmetric teleparallel equivalent extends to false-vacuum decay at leading semiclassical order. For a real scalar field minimally coupled to gravity, we compare the tunnelling exponent $B=S_{\rm b}-S_{\rm fv}$, which determines the exponential suppression of the decay rate, $\Gamma\sim e^{-B}$, with the Euclidean bounce, $S_{\rm b}$,  and false-vacuum actions, $S_{\rm fv}$. The three formulations admit the same $O(4)$-symmetric Coleman-De Luccia bounce, but the boundary terms relating their actions can contribute to $B$. We explicitly evaluate the torsion and non-metricity boundary contributions at the regular origin, $r=0$, and at the asymptotic endpoint, $r\to\infty$. For regular configurations with consistently matched bounce and false-vacuum boundary data, these contributions vanish in both TEGR and STEGR. We therefore obtain $B_{\rm GR}=B_{\rm TEGR}=B_{\rm STEGR}$, establishing that the three formulations predict the same leading semiclassical suppression of false-vacuum decay within the class of configurations and boundary conditions considered.
\end{abstract}

\maketitle

\section{Introduction}\label{sec:intro}  

One of the main goals of modern theoretical and experimental particle physics
is the search for New Physics (NP) beyond the Standard Model (BSM), even
though direct experimental searches have so far not revealed any clear evidence
for it. In this context, the stability analysis
\cite{Krasnikov:1978pu,Cabibbo:1979ay,Hung:1979dn,Politzer:1978ic,Anselm:1980mj,turner,rees,sher,Lindner:1985uk,sherrep,Lindner:1988ww,Ford:1992mv,Sher:1993mf,Altarelli:1994rb,quiro,isido,Espinosa:2007qp,lee}
of the electroweak (EW) vacuum plays a crucial role in constraining and
assessing the viability of possible BSM scenarios. Early studies were mainly
focused on establishing bounds on the Higgs boson and top quark masses. These
bounds were obtained either by requiring the Higgs effective potential
\(V(\phi)\) not to develop values lower than the electroweak minimum \(v\), so
that the latter is the stable, true vacuum of the theory \cite{Altarelli:1994rb,Sher:1993mf,Ford:1992mv,Lindner:1988ww,Lindner:1985uk,Cabibbo:1979ay,Krasnikov:1978pu,Hung:1979dn,Politzer:1978ic,Anselm:1980mj},
or by allowing for the possibility that our Universe resides in a metastable,
false vacuum state, namely a local minimum of \(V(\phi)\) with a lifetime longer
than the age of the Universe\cite{turner,rees,sher,isido}. The discovery of
the Higgs boson in 2012
\cite{higgsmass,ATLAS:2014wva} has significantly renewed interest in the
stability problem, motivating refined analyses of the stability properties of
the EW vacuum, with particular attention to the question of whether it is
absolutely stable or metastable
\cite{isidue,millington1,grinstein,millington2}. Further studies have also
focused on investigating the cosmological implications of vacuum stability
during and after inflation
\cite{raja1,khan,raja2,kearney,goldberg,macdonald,ema1,ema2,okada,urbanowski1,urbanowski2},
as well as on assessing the impact that different NP scenarios may have on the
stability of the EW vacuum
\cite{NNLO,isiuno,isidue,bu,haba,ferreira,chaka1,Branchina:2019tyy,Branchina:2018xdh,Bentivegna:2017qry,Branchina:2016bws,Branchina:2018qlf,Branchina:2015nda,Branchina:2014rva,Branchina:2014usa,Branchina:2013jra}.

The theoretical description of metastable vacuum decay has its roots in the
work of Bender and collaborators \cite{bender},  who studied tunneling in quantum mechanical systems with several degrees of freedom using the WKB approximation. In quantum field theory (QFT), early
investigations of metastable vacua in scalar field theories were presented
in \cite{Lee:1974ma}, and subsequently in \cite{Kobzarev:1974cp}, where the
tunneling process was described in close analogy with nucleation phenomena in
statistical physics, such as the boiling of a superheated fluid. In this
picture, the decay of the metastable phase proceeds through the nucleation of a
bubble of the lower-energy state, namely the true vacuum, within the surrounding
false vacuum, triggered by quantum fluctuations. Along these lines, subsequent
works further developed the study of metastable vacuum decay in specific
field theoretical models. The semiclassical interpretation of
vacuum lifetimes and the nature of metastable states was further analyzed
in \cite{Stone:1975bd,Stone:1976qh}, and the phenomenological consequences of
vacuum stability in the Goldstone and the
Weinberg-Salam models were explored in \cite{Frampton:1976pb}.

The intuitive picture of bubble nucleation was given a systematic
semiclassical formulation by Coleman and Callan \cite{coleman}, who provided
the first complete and Lorentz-invariant description of false vacuum decay in
QFT, as pointed out in \cite{sherrep}.

In their approach, the decay of a metastable vacuum is described in terms of a saddle-point approximation to the Euclidean path integral. The relevant saddle point is the so-called \textit{bounce}, a non-trivial solution of the Euclidean equations of motion, satisfying the appropriate false vacuum boundary conditions. At leading semiclassical order, the decay rate $\Gamma$ is written in terms of the false vacuum-subtracted on-shell Euclidean action, $\Gamma \simeq D e^{-B}, B = S_{\rm b} - S_{\rm fv},$.
Here, $\hbar=1$ , $S_{\rm b}$ and \(S_{\rm fv}\) denote the
Euclidean actions evaluated on the bounce and false vacuum configurations, respectively, while $D$ is the prefactor determined by fluctuations around these configurations.

The setup
considered by Coleman and Callan consists of a scalar field theory whose
potential \(V(\phi)\) has a local minimum, corresponding to the false
vacuum at \(\phi_{\rm fv}\), and a deeper minimum, corresponding to the true
vacuum at \(\phi_{\rm tv}\). In particular, they studied the regime in which
the energy density difference,
\(V(\phi_{\rm fv}) - V(\phi_{\rm tv})\), is much smaller than the height of the
potential barrier separating the two vacua,
\(V(\phi_{\rm top}) - V(\phi_{\rm fv})\), where \(\phi_{\rm top}\) denotes the
field value at the top of the barrier. In this limit, the two phases (true/false vacuum regions) are separated by a \qm{thin wall}. In this regime, it is possible to resort to
an analytic treatment of the decay process within the so-called thin-wall
approximation \cite{coleman}. This semiclassical formalism was later extended by Linde to
finite-temperature field theory, enabling the study of vacuum decay in thermal
environments, such as those relevant for the early Universe \cite{Linde:1977mm}.

The inclusion of gravity in the vacuum stability analysis was pioneered by
Coleman and De Luccia in \cite{cdl}, where the tunneling process was studied in
an \(O(4)\)-symmetric (maximally symmetric) Euclidean curved background. In particular, the decay of a
Minkowski false vacuum into an Anti-de~Sitter (AdS) true vacuum provides a
clean setting to assess the role of gravitational backreaction in vacuum
stability. Within the thin-wall approximation, it was found that when the
Schwarzschild radius associated with the true vacuum bubble is much smaller
than the bubble size, so that gravitational effects are weak, the decay rate
closely reproduces the Minkowski spacetime result. Conversely, when the
Schwarzschild radius becomes comparable to the bubble size, gravitational
effects become strong and gravity can stabilize the false vacuum, preventing
the nucleation of a true vacuum bubble.
Although these results are derived within the thin-wall approximation, the
underlying physical mechanism is expected to be more general. In an
\(O(4)\)-symmetric gravitational background, the geometry involved in
the decay process is described in terms of a radial function \(\rho(r)\),
which determines the size of the three-spheres along the Euclidean radial
coordinate \(r\). Compared to case where gravity is not taken into account, for \(\rho(r)=r\),
the gravitational backreaction modifies the geometry associated with
the bubble configuration, effectively \qm{shrinking} the volume of the
true vacuum region. As a consequence, the negative contribution to the
(Euclidean) action coming from the lower vacuum energy inside the bubble is
reduced in magnitude. The action governing the decay process is
therefore increased, which makes the decay rate exponentially more suppressed.

In the Coleman-De~Luccia analysis, the gravitational interaction is described
within the Riemannian framework of General Relativity (GR), where gravity is
encoded in the curvature of the Levi-Civita connection. Introduced by Einstein
in 1915, GR is based on general covariance, the Equivalence Principle (EP) and
the Principle of Causality \cite{Misner:1973prb}. In this framework the metric \(g_{\mu\nu}\) fixes
lengths, angles and causal relations, while the linear connection
\(\Gamma^\alpha{}_{\mu\nu}\) defines parallel transport and free-fall
trajectories. Although these two fields can be a priori independent,
in GR the EP selects the Levi-Civita connection (constructed uniquely from the
metric) as the connection defining locally inertial observers \cite{Schiff:1960ggq,Coley82}.

Nevertheless, both theoretical and observational considerations motivate the
study of extensions of GR, as well as of alternative geometric formulations of
the gravitational interaction \cite{Bonola1955,Bennett2011,Sotiriou:2008rp,Capozziello:2025hyw}. From a
theoretical perspective, GR has not yet been fully reconciled with quantum
field theory within a complete theory of quantum gravity. Moreover, its
classical dynamics predicts spacetime singularities, signaling the limits of
the classical description and motivating the search for more fundamental
frameworks \cite{Hooft1974,Obukhov_2017,Giulini2003,Rovelli2004}. From an observational
perspective, although the standard \(\Lambda\)CDM model successfully accounts
for a broad range of data, it relies on ingredients whose microphysical origin
is still uncertain \cite{Perivolaropoulos:2021jda,DESI:2024mwx}, such as dark matter
\cite{Mayet:2016zxu,Capozziello:2017rvz,Arun_2017}, dark energy
\cite{Carroll:2000fy,Planck:2018vyg}, and the mechanism responsible for
inflation \cite{Guth:2005zr}. In addition, persistent discrepancies in
precision cosmology, most notably the \(H_0\) and \(\sigma_8\) tensions \cite{Aloni:2021eaq,DiValentino:2021izs,Pantos:2026koc}, further motivate
the exploration of the gravitational sector. 

For these reasons, in the last century many attempts have been made to go
beyond the standard Riemannian formulation of GR by modifying and/or extending
the underlying geometric framework. 
A natural geometric generalization is to relax the assumption that the
affine connection coincides with the Levi-Civita connection. In a
metric-affine framework, the connection is independent of the metric and
need be neither symmetric nor metric-compatible. As a result, the geometry of the manifold is characterized
not only by curvature, but also by torsion and non-metricity. These three
tensors are intrinsic properties of the affine connection and describe
different aspects of parallel transport. They define the general framework of
metric-affine geometry \cite{Bahamonde:2015zma,Coley:2019zld}. Teleparallel theories form a class of metric-affine theories of gravity (MAGs) in which the linear affine connection is flat, so that it has vanishing curvature.

The present work focuses on two teleparallel formulations that are
classically equivalent to GR. In the metric teleparallel equivalent of GR
(TEGR), the connection is flat and metric-compatible, and gravity is
represented by torsion. In the symmetric teleparallel equivalent of GR (STEGR), the connection is flat and torsion-free, and gravity is represented by non-metricity. The gravitational Lagrangians of GR, TEGR, and STEGR differ
by total derivatives and yield the same bulk field equations. Together, these constitute the {\it Geometric Trinity of Gravity} \cite{BeltranJimenez:2019esp,Capozziello:2022zzh}. This framework gives rise to several cosmological applications 
\cite{Battista:2026gvo}. From the theoretical point of view, it can be seen that such a structure naturally emerges from a pre-geometric picture of gravity \cite{Capozziello:2026pys}.

The trinity offers complementary geometric and gauge-theoretic perspectives
on the gravitational interaction \cite{Pereira:2013qza,Brezina:2025dbc}. Its
relation to the universality of free fall and the Equivalence Principle (EP)
has been examined in
\cite{Altschul:2014lua,Tino:2020nla,Alonso:2022oot}; the role of the EP and
Schiff's conjecture in teleparallel formulations has likewise been discussed
\cite{Capozziello:2024ijv,Schiff:1960ggq,Coley82}. More general metric-affine
settings also admit test particles with intrinsic hypermomentum, whose
equations of motion have been investigated in
\cite{Iosifidis:2023eom,Iosifidis(2026),Ferrara:2026bzy}. Within the Geometric Trinity, however,
curvature, torsion, and non-metricity provide alternative descriptions of
the same classical gravitational degrees of freedom (DoFs).

This classical equivalence raises the question of how far it extends to quantum phenomena. False vacuum decay offers a way to address this question
at leading semiclassical order: its tunneling exponent depends on the value of the Euclidean action, rather than solely on the field equations. 

In this work, we consider a single real scalar field minimally coupled to gravity and
formulate the Coleman-De Luccia decay problem in TEGR and STEGR at the leading order of the semiclassical expansion. 
With the
same regularity and false vacuum boundary conditions as in GR, the three
formulations admit the same scalar-field and metric bounce configurations.
This agreement does not, by itself, establish equality of their tunneling
exponents. A boundary term whose variation vanishes under the admissible
boundary conditions can still contribute to the on-shell action. The
torsion and non-metricity currents entering the teleparallel actions must
therefore be evaluated explicitly.

The calculation will be performed according the Coleman-De Luccia analysis \cite{cdl}.

We evaluate the boundary contributions on both the bounce and false vacuum configurations, examining the endpoints of the Euclidean radial domain, \(r\to0\) and \(r\to\infty\). Under the boundary conditions and for the teleparallel configurations considered here, these contributions vanish in both TEGR and STEGR. Comparing the resulting on-shell actions with their GR counterparts therefore yields
$B_{\rm TEGR}=B_{\rm STEGR}=B_{\rm GR}.$

This equality does not follow from the coincidence of the bulk bounce configurations alone. The tunneling exponent  $B=S_{\rm b}-S_{\rm fv}$ depends on the values of the on-shell Euclidean actions. 
A boundary term can contribute to these values even when its variation vanishes under the prescribed boundary conditions. The boundary contributions associated with the torsion and non-metricity currents must therefore be evaluated explicitly.\footnote{Their evaluation requires careful consideration of the regularity, asymptotic behavior, and gauge sector of the teleparallel variables.}

The three formulations consequently predict the same exponential suppression of false vacuum decay within the setting studied here, establishing their equivalence for this observable at leading exponential order. A comparison of the fluctuation prefactors and an investigation of equivalence beyond this order are left for future work.

The paper is organized as follows. In Section\ \ref{sec:MAGs}, we review the geometrical framework underlying TEGR and STEGR, emphasizing their relation with GR within the Geometric Trinity. In Section\ \ref{sec:thback}, we recall the semiclassical calculation of false vacuum decay in GR and the role of gravitational backreaction in the Coleman-De Luccia construction. We then reformulate the decay problem in STEGR, Section\ \ref{vacdecaySTEGR}, and in TEGR, Section\ \ref{vacdecayTEGR}. Finally, we compare the three descriptions to
establish the equivalence of their tunneling exponents.

\section{Geometrical formulation of TEGR and STEGR}\label{sec:MAGs}

A natural way of extending the geometrical framework of Einstein gravity is to relax the assumption that the affine connection is necessarily the Levi-Civita one.
A metric-affine theory is defined by the triplet $\{\mathcal{M},g_{\mu\nu},\Gamma^{\rho}_{\ \mu\nu}\}$, where $\mathcal{M}$ is a four-dimensional spacetime manifold, $g_{\mu\nu}$ is a rank-two symmetric tensor (with 10 independent components), and $\Gamma^{\rho}_{\ \mu\nu}$ is the affine connection (endowed with 64 independent components). In general, the metric and the affine connection do not have to be related; in fact, the former describes the \emph{causal structure} and the latter defines the \emph{parallel transport}. However, the two structures coincide if the theory is built on the EP, since it relates the affine connection to derivatives of the metric, as in GR \cite{Will1993, Faraoni2010}. 

The covariant derivative $\nabla$ acts on a generic $(1,1)$ tensor in the following way 
\begin{equation}
\nabla_\mu A^{\alpha}_{\ \beta}:=\partial_\mu A^{\alpha}_{\ \beta}- \Gamma^{\rho}_{\ \beta\mu}A^{\alpha}_{\ \rho}+ \Gamma^{\alpha}_{\ \rho\mu}A^{\rho}_{\ \beta},
\end{equation}
and the most general linear connection can always be decomposed as
\begin{equation}\label{eq: general affine connection}
    \Gamma^{\rho}{}_{\mu\nu} = \overset{\circ}{\Gamma}{}^{\rho}{}_{\mu\nu} + N^{\rho}{}_{\mu\nu},
\end{equation}
where, next to the Levi-Civita connection
\begin{equation}
\gr{\Gamma}{}^{\rho}{}_{\mu\nu} = \dfrac{1}{2} g^{\rho\sigma}( \partial_{\mu} g_{\nu\sigma} + \partial_{\nu} g_{\sigma\mu} - \partial_{\sigma}g_{\mu\nu}),
\end{equation}
we define the tensor $N^\rho{}_{\mu\nu}$ as the difference between the affine connections $\Gamma^{\rho}{}_{\mu\nu}$ and $\gr{\Gamma}{}^{\rho}{}_{\mu\nu}$.
In this framework, $N^\rho{}_{\mu\nu}$ can be uniquely decomposed as:
\begin{equation}
    N^{\rho}{}_{\mu\nu} = K^{\rho}{}_{\mu\nu} + L^{\rho}{}_{\mu\nu},
\end{equation}
where $K^{\rho}_{\ \mu\nu}$ is the contortion tensor and $L^{\rho}_{\ \mu\nu}$ is the disformation tensor, whose explicit expressions are:
\begin{subequations}
\begin{align}
    K^{\rho}{}_{\mu\nu} &:= \dfrac{1}{2} g^{\rho \alpha}( -T_{\mu\alpha\nu} - T_{\nu\alpha\mu} + T_{\alpha\mu\nu} ) = - K_{\mu\ \nu}^{\ \rho} \label{eq: contorsion}, \\
    L^{\rho}{}_{\mu\nu} &:= \dfrac{1}{2} g^{\rho\alpha} ( -Q_{\mu\alpha\nu} - Q_{\nu\alpha\mu} + Q_{\alpha\mu\nu} ) = L^{\rho}{}_{\nu\mu} \label{eq: disformation}.
\end{align}
\end{subequations}
The Levi-Civita connection and the disformation tensor are symmetric in their last two indices, and the contortion tensor parametrizes the antisymmetric part of the connection through the torsion tensor
\begin{equation}\label{eq: torsion}
    T^{\rho}{}_{\nu\mu} = \Gamma^{\rho}{}_{\mu\nu} - \Gamma^{\rho}{}_{\nu\mu}.
\end{equation}
On the other hand, the disformation tensor is written in terms of the non-metricity tensor, which is given by 
\begin{equation}\label{eq: non-metricity}
    Q_{\rho\mu\nu} = \nabla_{\rho} g_{\mu\nu} = \partial_{\rho} g_{\mu\nu} - \Gamma^{\alpha}{}_{\mu\rho}\, g_{\alpha\nu} - \Gamma^{\alpha}{}_{\nu\rho}\, g_{\mu\alpha},
\end{equation}
and represents the deviation of the connection from metric-compatibility. The last property of the general linear connection is the curvature tensor, which reads
\begin{equation}\label{eq: curvaure}
    R^{\sigma}{}_{\rho\mu\nu} =  \partial_{\mu} \Gamma^{\sigma}{}_{\rho\nu} -  \partial_{\nu} \Gamma^{\sigma}{}_{\rho\mu} + \Gamma^{\alpha}{}_{\rho\nu} \Gamma^{\sigma}{}_{\alpha\mu} - \Gamma^{\alpha}{}_{\rho\mu}\Gamma^{\sigma}{}_{\alpha\nu}.
\end{equation}

These three main geometric objects are related to the dynamics and differently affect the parallel transport of a vector on a manifold:
\begin{itemize}
    \item \emph{curvature} causes a non-null angle when a vector is parallel transported along a closed curve on a non-flat background; 
    \item \emph{torsion} entails a rotational geometry, where the parallel transport of two vectors is antyisymmetric by exchanging the transported vectors and the direction of transport. This property results in the non-closure of parallelograms; 
    \item \emph{non-metricity} alters the length and the angles of the vectors along the transport. 
\end{itemize}
In metric-affine framework, the following geometric relation holds:
\begin{equation}\label{eq: general R}
    R = \gr{R}+T+\hat{B}-Q+B + T_{\rho\mu\nu} Q^{\mu\nu\rho} 
- T_\mu Q^\mu 
+ T_\mu \tilde{Q}^\mu ,
\end{equation}
where $R$ is the general Ricci scalar, defined with respect to the general linear affine connection \eqref{eq: general affine connection} and $\gr{R}$ is the GR Ricci scalar. Moreover
\begin{subequations}
\begin{align}    Q_\alpha&:=Q_{\alpha\lambda}{}^\lambda \label{eq:Qa}\\
\tilde{Q}_\alpha&:=Q^\lambda_{\ \lambda\alpha}\label{eq:tildeQa}
\end{align}
\end{subequations}
represent two independent traces of the non-metricity tensor, while
\begin{equation}\label{eq:Ta}
   T^{\alpha\mu}{}_{ \alpha}:=T^\mu 
\end{equation}
is the \emph{torsion vector} \cite{Capozziello:2022zzh,Bahamonde:2021gfp}. Finally, $Q$ is the non-metricity scalar with its related boundary term $B$:
\begin{subequations}
    \begin{align}\label{eq: Q}
        Q &:= \frac{1}{4}(Q^{\alpha\rho\nu}Q_{\alpha\rho\nu}) - \frac{1}{2}(Q^{\alpha\rho\nu}Q_{\rho\nu\alpha})-\frac{1}{4}Q^{\alpha}Q_{\alpha} + \frac{1}{2} \tilde{Q}^{\alpha}Q_{\alpha}, \\
        B&:=\gr{\nabla}_{\alpha}(Q^{\alpha}-\tilde{Q}^{\alpha})\label{eq: boundaryQ},
    \end{align}
\end{subequations}
and $T$ is the torsion scalar, with its related boundary term $\hat{B}$:
\begin{subequations}
\begin{align}\label{eq:general_torsion}
 T&:=-\frac{1}{4} T_{\alpha \mu \nu} T^{\alpha \mu \nu}-\frac{1}{2} T_{\alpha \mu \nu} T^{\mu \alpha \nu}+T_{\alpha} T^{\alpha},\\
 \label{eq: boundaryT}
 \hat{B}&:= 2\gr{\nabla}_\alpha T^{\alpha} .
 \end{align}
\end{subequations}

Among the possible metric-affine gravity theories, Riemannian and teleparallel models are particularly interesting. 
GR is an example of Riemannian geometry, whereas the so-called metric teleparallel equivalent of GR (TEGR) and symmetric teleparallel equivalent of GR (STEGR) are examples of teleparallel geometries. These three theories form the so-called \emph{Geometric Trinity of Gravity} \cite{BeltranJimenez:2019esp}. 
The last two are teleparallel theories because they are curvature-less. In fact, TEGR and STEGR are based on the concept of \emph{Fernparallelismus} or \emph{parallelism at distance}: the parallel transport of vectors becomes independent of the path. 
Thus, the linear affine connection $\Gamma^\alpha{}_{\mu\nu}$ is forced to be flat:
\begin{equation}\label{eq: flat_curvature}
    R^{\alpha}_{\ \beta\mu\nu}=0,
\end{equation}
which means that we can find a coordinate transformation $\xi^{\lambda}\to x^{\lambda}$ such that:
\begin{equation}\label{eq: transf coordinate CG}
    \Gamma^{\rho}{}_{\mu\nu}(x^\lambda) = \frac{\partial x^{\rho}}{\partial \xi^{\gamma}} \frac{\partial \xi^{\alpha}}{\partial x^{\mu}} \frac{\partial \xi^\beta}{\partial x^{\nu}} \Gamma^{\gamma}{}_{\alpha \beta}(\xi^{\lambda}) + \frac{\partial^{2} \xi^{\alpha}}{\partial x^{\mu} \partial x^{\nu}} \frac{\partial x^{\rho}}{\partial \xi^{\alpha}}.
\end{equation}
The flatness condition restricts the connection to be purely inertial and Eq.\,\eqref{eq: transf coordinate CG} can be parameterized by the general element $\Lambda^{\alpha}_{\phantom{\alpha}\beta}$ of $GL(4,\mathbb{R})$ as follows:
\begin{equation}\label{eq: gl4}
\Gamma^{\alpha}_{\phantom{\alpha}\mu\nu}=(\Lambda^{-1})^{\alpha}_{\phantom{\alpha}\lambda}\partial_{\mu}\Lambda^{\lambda}_{\phantom{\lambda}\nu} 
\end{equation}
because of the independence of the details of the transformation \cite{BeltranJimenez:2019esp, BeltranJimenez:2017tkd}.

A fundamental property of both TEGR and STEGR is that torsion and non-metricity respectively replace the curvature for dynamics, providing the same descriptions of the gravitational interaction. In GR, the geometric curvature models the gravitational force, since geodesics coincide with the free-falling test particle's trajectories. On the other hand, in TEGR, the gravitational interaction emerges through the torsion tensor and acts as a (gauge) force. This is the reason why, in the teleparallel framework, the concept of geodesics is replaced by force equations, analogously to what happens in electrodynamics where  the Lorentz force is present. STEGR shares properties similar to those of TEGR. In this theory, one requires that curvature and torsion are both zero, and gravitational dynamics is attributed to the non-metricity tensor. 

GR is described in terms of the metric $g_{\mu\nu}$; TEGR in terms of the tetrads $e^a_{\ \mu}$ (accounting for the dynamical description of gravity) and spin connection $\omega^a_{\ b\mu}$ (flat connection outlining inertial effects); STEGR embodies the Palatini idea to separate the metric $g_{\mu\nu}$ and affine connection $\Gamma^\mu_{\ \alpha\beta}$, considering them as two different dynamical structures. In the teleparallel formulations, the EP can be recovered in such theories,  even if it does not lie at their foundation \cite{Pereira:2013qza, Capozziello:2022zzh, Capozziello:2024ijv, Mancini:2025asp}. This fact is extremely relevant because, if the EP  were shown to be violated at some fundamental level, the final theory of gravitation could be non-metric \cite{Capozziello:2024ijv}. 

These equivalent pictures define alternative ways of representing the  gravitational field, accounting for the same DoFs, related to specific geometric   invariants: the Ricci scalar $R$, the torsion scalar $T$, and the non-metricity scalar $Q$. In this perspective, GR, TEGR, and STEGR give rise to the Geometric Trinity of Gravity. 

\subsection{Symmetric Teleparallel Equivalent of GR}
\label{Sec: STEGR}

Among teleparallel models, \textit{symmetric teleparallel theories} describe the gravitational interaction in terms of the non-metricity tensor $Q_{\alpha\mu\nu}$, defined in Eq.\ \eqref{eq: non-metricity}. In this framework, the affine connection is flat, satisfying the condition given in Eq.\ \eqref{eq: flat_curvature}, and torsion-free:
\begin{equation}\label{eq:torspost}
T^{\alpha}{}_{\mu\nu}=0.
\end{equation}
In a coordinate basis, the latter condition implies that the connection coefficients are symmetric in their two lower indices.
Hence, the independent geometrical object that encodes gravity is the 
non-metricity tensor \cite{BeltranJimenez:2019esp,Hohmann:2019nat,Capozziello:2022zzh}.
In this framework, the most general even-parity quadratic form of the
non-metricity tensor can be written as
\begin{equation}\label{eq: general_nonmetr}
    Q_{\rm gen} := c_1 Q_{\alpha \mu \nu} Q^{\alpha \mu \nu}+c_2 Q_{\alpha \mu\nu} Q^{\mu \alpha \nu}+c_3 Q_{\alpha} Q^{\alpha}+c_4 \tilde{Q}_{\alpha} \tilde{Q}^{\alpha}+c_5 Q_{\alpha} \tilde Q^{\alpha},
\end{equation}
where \(c_i\) are free constant parameters. This gives rise to the \emph{five-parameter family of quadratic theories} or the so-called \emph{Newer GR}, see
Ref.\ \cite{BeltranJimenez:2017tkd,Ferrara:2026hbq} for more details. In particular, STEGR is obtained by choosing
\(c_1=c_3=1/4\), \(c_2=c_5=1/2\), and \(c_4=0\), which gives
the non-metricity scalar appearing in Eq.\ \eqref{eq: Q},
\begin{equation}\label{eq: non-metricity scalar}
    Q := P^{\alpha\mu\nu}Q_{\alpha\mu\nu}=\frac{1}{4}\left( Q_{\alpha} Q^{\alpha}-Q_{\alpha \mu \nu} Q^{\alpha \mu\nu}\right)+\frac{1}{2}\left( Q_{\alpha \mu \nu} Q^{\mu \alpha \nu}-Q_{\alpha} \tilde{Q}^{\alpha}\right),    
\end{equation}
where the tensor \(P^{\alpha\mu\nu}\) is the \emph{superpotential}, or the
\emph{non-metricity conjugate},
\begin{equation}\label{eq:superpotSTEGR}
    P^\alpha{}_{\mu\nu}:=\frac{1}{2\sqrt{-g}}\frac{\partial(\sqrt{-g}Q)}{\partial Q_\alpha^{\ \mu\nu}}=\frac{1}{4}Q^\alpha_{\ \mu\nu}-\frac{1}{4}Q_{(\mu}{}^\alpha_{\ \nu)}-\frac{1}{4}g_{\mu\nu}Q^{\alpha\beta}{}_\beta+\frac{1}{4}\left[Q_\beta^{\ \beta\alpha}g_{\mu\nu}+\frac{1}{2}\delta^\alpha_{(\mu}Q_{\nu)}{}^\beta_{\ \beta}\right].
\end{equation}
With the non-metricity scalar chosen as in Eq.\,\eqref{eq: non-metricity scalar}, the general metric-affine identity \eqref{eq: general R} applies and, using the flatness
condition \eqref{eq: flat_curvature} and the torsionless condition
\eqref{eq:torspost}, it reduces to
\begin{equation}\label{eq:GR_STEGR}
    \gr{R}= Q-B= Q -\gr{\nabla}_\mu(Q^\mu-\tilde{Q}^\mu).    
\end{equation}
This identity shows that the STEGR scalar \(Q\) in Eq.\,\eqref{eq: non-metricity scalar} reproduces the
Einstein-Hilbert (EH) Lagrangian up to the non-metricity divergence term.

As a consequence of Eq.\ \eqref{eq:GR_STEGR}, we can define the STEGR gravitational action as
\begin{equation}\label{eq:STEGR_Lagrangian}
S_{_{\rm STEGR}}[g,\Gamma]:=\int_{\mathcal M} {\rm d}^{4}x\ \sqrt{-g} \frac{c^4}{16\pi G}Q,
\end{equation}
which reproduces the same bulk
gravitational dynamics as GR. The coupling to matter within the framework of symmetric teleparallel theories can be performed
following the usual minimal coupling prescription (for simplicity, we call the total action $S_{_{\rm STEGR}}$ as for the pure gravity case):
\begin{equation}\label{eq: STEGR action}
    S_{_{\rm STEGR}}[g,\Gamma,\Psi] := \int_{\mathcal M} {\rm d}^{4}x\ \sqrt{-g}\left[ \frac{c^4}{16\pi G}Q + \mathcal{L}_{\rm m} \right],
\end{equation}
where $\Psi$ generically denotes matter fields.
With this prescription, matter fields couple to the metric as in GR and do
not directly probe the independent affine connection
\cite{BeltranJimenez:2019esp}. For the scalar field considered below, one
therefore has \(\nabla_\mu \phi=\partial_\mu\phi\), as in the usual
formulation of GR.

The metric field equations are obtained by varying the action with respect to
\(g^{\mu\nu}\) (\(\kappa=8\pi G\)):
\begin{align}\label{geqm}
\frac{2}{\sqrt{-g}}\nabla_{\alpha}\left(\sqrt{-g} P^\alpha{}_{\mu\nu}\right)-\frac 12 g_{\mu\nu}\,Q+\left( P_{\mu\alpha \beta}Q_\nu{}^{\alpha \beta}-2 Q_{\alpha \beta \mu}P^{\alpha \beta}{}_\nu\right)=\kappa \mathfrak{T}_{\mu\nu}\,,
\end{align}
where \(\mathfrak{T}_{\mu\nu}\) is the stress-energy tensor.
The left-hand side of Eq.\ \eqref{geqm} is equivalent to the Einstein tensor $ \gr{G}_{\mu\nu}$,
so that:
\begin{equation}
    \gr{G}_{\mu\nu}:=\frac{2}{\sqrt{-g}} \nabla_{\alpha}\left(\sqrt{-g} P^{\alpha}{}_{\mu \nu}\right)
   -\frac{1}{\sqrt{-g}}q_{\mu \nu}+\frac{1}{2} g_{\mu \nu}Q.
\end{equation}
Moreover, the connection field equations, obtained by varying with respect to
\(\Gamma^\alpha{}_{\mu\nu}\), are
\begin{align}\label{connectioneqm}
\nabla_{\mu}\nabla_\nu\left(\sqrt{-g} P^{\mu\nu}{}_{\alpha}\right)=0\,,
\end{align}
and they are identically satisfied, according to the Bianchi identity, when
the dynamical equivalence to GR holds, Eq.\ \eqref{geqm}. Therefore, in STEGR
the dynamics of the theory is entirely determined by the Einstein field
equations.

At the level of the action \eqref{eq: STEGR action}, the dynamical equivalence follows from the Levi-Civita
divergence identity
\begin{equation}
 \gr{\nabla}_\mu\left(Q^\mu-\tilde{Q}^\mu\right)
 \equiv
 \frac{1}{\sqrt{|g|}}\,
 \partial_\mu
 \left[
 \sqrt{|g|}\left(Q^\mu-\tilde{Q}^\mu\right)
 \right].
\end{equation}
More explicitly, for any vector \(V^\mu\),
\begin{equation}\label{eq:divergencetoboundary}
  \int_{\mathcal M} d^4x\,\sqrt{|g|}\,
  \gr{\nabla}_\mu V^\mu
  =
  \int_{\partial \mathcal M} d^3x\,\sqrt{|h|}\,
  n_\mu V^\mu,
\end{equation}
where \(n^\mu\) is the unit normal to the boundary \(\partial M\), and
\(h_{ij}\) is the induced metric on it. Thus, when the appropriate boundary
data are held fixed, the variation of this total derivative gives only a
boundary contribution and does not modify the equations of motion in the
interior of \(M\). This is the sense in which STEGR and GR are dynamically
equivalent at the bulk classical level. 
As stressed in the Introduction, however, this variational statement does
not automatically imply that the boundary term itself vanishes when evaluated on a solution.
This is particularly relevant for the vacuum decay problem considered in this work,
where the bounce and the false vacuum solutions are non-compact, which might give rise to finite contributions to the action from the endpoints.

Finally, let us recall the pure-gauge form of the STEGR connection. When the
torsionless condition is implemented, the connection defined with respect to
the \(\Lambda\) matrices of Eq.\ \eqref{eq: gl4} satisfies
\begin{equation}
\partial_{\mu}\Lambda^{\lambda}_{\phantom{\lambda}\nu}
-
\partial_{\nu}\Lambda^{\lambda}_{\phantom{\lambda}\mu}
=
0.
\end{equation}
Introducing functions of the \(x\) coordinates
\(\xi^{a}=\xi^{a}(x^{\nu})\)
\cite{Capozziello:2021pcg,DAmbrosio:2021zpm} and defining
\begin{equation}\label{eq: scalarconnection}
	\Lambda^{\mu}_{\phantom{\mu}\nu}=\partial_{\nu}\xi^{\mu}, 
\end{equation}
we have
\begin{equation}\label{eq: coincident gauge}
\Gamma^{\alpha}_{\phantom{\alpha}\mu\nu}=\frac{\partial x^{\alpha}}{\partial \xi^{a}}\partial_{\mu}\frac{\partial \xi^{a}}{\partial x^{\nu}}.
\end{equation}
Thus, the connection ascribed to symmetric teleparallel theories can be
trivialized by the choice \(\xi^{\mu}=x^{\mu}\), or by the more general
relation \(\xi^{\mu}=M^{\mu}_{\phantom{\mu}\nu}x^{\nu}+\xi_c^{\mu}\), where
the matrix \(M\) is invertible and its components are constants
\cite{DAmbrosio:2021zpm,Jarv:2023sbp}.

This particular coordinate choice is called \textit{coincident gauge}, and it
can be interpreted as the gauge in which the origin of the tangent space,
parameterized by the scalar functions \(\xi^\mu\), coincides with the
spacetime origin \cite{BeltranJimenez:2019esp,Capozziello:2022zzh}. Its
availability is guaranteed when imposing the curvatureless and torsionless
postulates \cite{Capozziello:2025hyw}.

\subsection{Tetrad Formalism}\label{sec: tetrad formalism}

Metric (or torsional) teleparallel gravity is obtained by assuming the metric compatibility. The theory is geometrically given only by the torsion tensor, and its framework is completely described by the tetrad \(e^a{}_\mu\), which encodes the gravitational field, and the spin
connection \(\omega^a{}_{b\mu}\), which accounts for inertial effects.

A tetrad field is a geometric construction, which allows to easily solder the tangent space to the spacetime manifold. Physically, tetrads are the standard laboratory-apparatus of the observer for carrying out the measurements in space and time. This geometric structure is always present, independently of any prior gravity-model assumption. 
The spacetime is the base space and the tangent vector space is the fiber bundle, attached at each point of it. In four dimensions, the spacetime manifold and the tangent bundle are soldered in such a way that the spacetime metric $g_{\mu\nu}$ and the Minkowski metric $\eta_{ab}$ are connected by the tetrad field $e^{a}{}_{\mu}$\,:
\begin{equation}\label{eq: metric}
g_{\mu\nu}=e^{a}_{\phantom{a}\mu}e^{b}_{\phantom{b}\nu}\,\eta_{ab}.
\end{equation}

On the fiber bundle, i.e. the Minkowski tangent space, the gauge transformations represent local translations; thus, we have to respect the invariance under both general coordinate transformations, performed in the spacetime, and local Lorentz transformations, performed in the tangent space.
In this framework, the inertial effects are represented by a purely inertial connection, the \textit{spin connection} or \textit{Lorentz connection}, which depends on local Lorentz transformations and defines different classes of frame. The equivalent inertial frames are related by a global Lorentz transformation, and the class of frames without inertial effects is characterized by a trivial inertial Lorentz connection. 
In all other classes of frame, the inertial spin connection is not-vanishing. A natural differentiable basis of $T_{p}\mathcal{M}$ is given by the sets of gradients
\begin{equation}
    \left\{\partial_{\mu}\right\}:=\left\{\frac{\partial}{\partial x^{\mu}}\right\}
\end{equation}
as well as on the cotangent space $T^*_{p}\mathcal{M}$ the covector basis is $\left\{dx^{\mu}\right\}$. Both $\left\{\partial_{\mu}\right\}$ and $\left\{dx^{\mu}\right\}$ satisfy the orthonormality conditions:
\begin{subequations}
\begin{align}\label{eq: ortho condition}
dx^{\mu}\partial_{\nu}&=\delta^{\mu}_{\nu}\\
    dx^a \partial_b&=\delta^a_b.\label{eq:ortho2}
\end{align}
\end{subequations}

The entire set of such bases represents a basis for the vectors on $T_p\mathcal{M}$, at each point $p\in\mathcal{M}$. Thus, on the
common domains they are defined, we can express each orthonormal vector and covector with respect to the other \cite{Golovnev:2017dox,Pereira:2012kd}. 
The frame $e^{a}_{\phantom{a}\mu}$ is such that respects the orthonormality condition given in Eqs. \eqref{eq: ortho condition} and \eqref{eq:ortho2}:
\begin{align}
&e^{a}_{\phantom{a}\nu}e_{a}^{\phantom{a}\mu}=\delta^{\mu}_{\nu}\label{eq: orthogonality relation}\\
	&e^{a}_{\phantom{a}\mu}e_{b}^{\phantom{b}\mu}=\delta_{b}^{a}.
\end{align}

As $g_{\mu\nu}$ varies from point to point on the manifold $\mathcal{M}$, the vierbein $e^{\ \mu}_a$ do the same. The determinant of Eq. \eqref{eq: metric}, gives $-g=e^2$, in which the determinant of $e^{\ \mu}_a$ is identified by $e$ and it is negative owed to the signature of $\eta_{ab}$.

A general linear frame is expressed by tetrads or vierbein:
\begin{align}\label{eq: Tetrad}
	e_{a}&=e_{a}^{\phantom{a}\mu}\partial_{\mu}\\
	e^{a}&=e^{a}_{\phantom{a}\mu}dx^{\mu}
    \label{eq: Tetradinv}
\end{align}
that satisfy the commutation relations:
\begin{equation}\label{eq: commutatori tetradi}
    [e_a, e_b ] := f^{c}{}_{ab} e_c.
\end{equation}
The quantities $f^{c}{}_{ab}$ are the structure coefficients or the anholonomy of frame:
\begin{equation}\label{eq: coeff struttura}
     f^{c}{}_{ab} = e^{\mu}_{a}e^{\nu}_{b}(\partial_{\nu}e^{c}_{\mu} - \partial_{\mu}e^{c}_{\nu}).
\end{equation}
The structure coefficients $f^{c}{}_{ab}$ are functions of the spacetime points and measure the not-closure of the parallelogram formed by the vectors $e_a$ and $e_b$ \cite{Capozziello:2022zzh}. If $f^{c}{}_{ab} \neq 0$, the tetrad basis is said to be \textit{anholonomic} or \textit{non-trivial} tetrads.
However, in a class of frames \(e'_a\), it is possible to set
\begin{equation}\label{eq: olonomia}
	f^{c}{}_{ab}=0,
\end{equation}
locally. This is the class of inertial frames, in which the tetrad basis is
holonomic. In fact, Eq.\ \eqref{eq: olonomia} means that \(e'^{a}\) is locally
an exact differential form and there exists a neighborhood around the point
\(p\) on which
\begin{equation}\label{eq:proper}
		e'^{a}=dx'^{a}
		=
		\partial_\mu x'^{a}\,dx^\mu
		\quad \Rightarrow \quad
		e'^a{}_\mu=\partial_\mu x'^a .
\end{equation}
In absence of gravitation, anholonomy is only caused by the inertial effects
associated with the chosen frame: the metric \(g_{\mu\nu}\) represents the
Minkowski metric \(\eta_{\mu\nu}\) in a general coordinate system. In absence
of inertial effects, the class of inertial frames is characterized by
vanishing structure coefficients, since all coordinate bases are holonomic.
This property is valid everywhere for frames belonging to this inertial class
\cite{Pereira:2013qza,Pereira:2012kd,Krssak:2015rqa,Krssak:2018ywd}. 
In the terminology of covariant teleparallel gravity, a tetrad that, in the
absence of gravitation, belongs to this inertial class and therefore does not
carry inertial effects is called a \emph{proper tetrad}.

The spin connection or \textit{Lorentz connection} $\omega_{\mu}$ is a 1-form acting in the Lorentz algebra:
\begin{equation}
\omega_{\mu}=\frac{1}{2}\omega^{ab}_{\phantom{ab}\mu}L_{ab},
\end{equation}
where $L_{ab}$ is a representation of the Lorentz generators and $\omega^{ab}{}_{\mu}$ are the spin connection coefficients. These connections are gauge potentials, which are introduced to define covariant derivatives under gauge transformations. They are related to the linear group $GL(4,\mathbb{R})$.
A tetrad field relates tangent space (or internal) tensors with those related to the spacetime (or external). It is possible to relate the connection 1-form $\omega$ and the linear affine connection:
\begin{equation}\label{eq: fock derivata}
    \Gamma^{\lambda}{}_{\mu \nu} = e^{\lambda}_{a}e^{b}_{\nu}\omega^{a}_{b\mu} + e^{\lambda}_{a}\partial_{\mu}e^{a}_{\nu} \equiv e^{\lambda}_{a}D_{\mu} e^{a}_{\nu},
\end{equation}
\begin{equation}\label{eq: cov derivata}
    \omega^{a}_{b\mu} = e^{a}_{\lambda} e^{\nu}_{b} \Gamma^{\lambda}{}_{\nu\mu} + e^{a}_{\sigma}\partial_{\mu}e^{\sigma}_{b} \equiv e^{a}_{\nu} \nabla_{\mu} e^{\nu}_{b}.
\end{equation} 
We stress that $\nabla_{\mu}$ is the covariant derivative defined with respect to the connection $\Gamma^{\lambda}{}_{\mu\nu}$,  acting  on external indices, and it can be defined for tensorial fields. On the other hand, the Fock-Ivanenko derivative $D_{\mu}$:
\begin{equation}
D_{\mu}=\partial_{\mu}-\frac{i}{2}\omega^{ab}_{\phantom{ab}\mu}L_{ab}
\end{equation}
acts on internal indices and can be defined for all tensorial and spinorial fields.
However, it is reasonable to assume $\nabla\equiv D$, because the same covariant derivative of a vector cannot change in terms of which type of basis one chooses \cite{Capozziello:2022zzh}. This is the so-called \emph{tetrad postulate}, which is valid for any affine connection, and it is encoded in both Eq. \eqref{eq: fock derivata} and Eq. \eqref{eq: cov derivata}: the total covariant derivative of the tetrad fields, expressed in terms of  connection for both internal and external indices, vanishes identically:
\begin{equation}\label{eq: tetrad postulate}
\nabla_{\mu}e^{a}_{\nu} = \partial_{\mu}e^{a}_{\nu}-\Gamma^{\lambda}{}_{\nu\mu}e^{a}_{\lambda}+\omega^{a}_{b\mu}e^{b}_{\nu} = 0.
\end{equation}

A local Lorentz transformation
$
x^a=\Lambda^a{}_{b}(x)x'^b
$
transforms the holonomic frame
$
	e'^a{}_\mu=\partial_\mu x'^a
$
into the locally rotated frame
\begin{equation}
	e^a{}_\mu
	=
	\Lambda^a{}_{b}(x)e'^b{}_\mu
	=
	\Lambda^a{}_{b}(x)\partial_\mu x'^b .
\end{equation}
Since the Lorentz matrix \(\Lambda^a{}_{b}(x)\) is spacetime dependent, the
ordinary derivative \(\partial_\mu x^a\) does not transform covariantly. The
rotated frame can instead be written in terms of the associated covariant
derivative as
\begin{equation}\label{eq:newframe}
	e^{a}_{\mu}
	=
	D_{\mu}x^{a}
	=
	\partial_{\mu}x^{a}
	+
	\omega^{a}{}_{b\mu}x^{b},
\end{equation}
where
\begin{equation}\label{eq: omega inerziale}
	\omega^{a}{}_{b\mu}
	=
	\Lambda^{a}{}_{c}
	\partial_{\mu}(\Lambda^{-1})^{c}{}_{b}
\end{equation}
is the Lorentz connection representing the inertial effects present in the
locally rotated frame, and \(D_\mu\) is the associated covariant derivative.
Under a local Lorentz transformation \(\Lambda^a{}_{b}(x)\), the spin
connection transforms as
\begin{equation}\label{eq: omega spuria}
	\omega^{a}{}_{b\mu}
	=
	\Lambda^{a}{}_{c}\,
	\omega'^{c}{}_{d\mu}\,
	(\Lambda^{-1})^{d}{}_{b}
	+
	\Lambda^{a}{}_{c}\,
	\partial_{\mu}(\Lambda^{-1})^{c}{}_{b}.
\end{equation}
In the right hand side of Eq.\ \eqref{eq: omega spuria}, we find two
contributions. The first term is the homogeneous transformation of the spin
connection already present in the original frame, while the second term is
the inhomogeneous contribution generated by the local Lorentz transformation.
Therefore, if one starts from an inertial frame in which the inertial spin
connection vanishes, \(\omega'^{c}{}_{d\mu}=0\), the transformed frame carries
the purely inertial connection given in Eq.\ \eqref{eq: omega inerziale}. This
connection represents the inertial effects associated with the rotated frame. Thus, starting from a proper tetrad and performing a local Lorentz
transformation \(\Lambda^{a}{}_{b}(x^\mu)\), one obtains a different class of
frames. Within each class, the infinitely many frames are related by global
Lorentz transformations, \(\Lambda^{a}{}_{b}=\mathrm{const}\). In inertial
frames, \(e'^{a}{}_{\mu}=\partial_{\mu}x'^{a}\), the inertial effects are
absent and the corresponding Lorentz connection vanishes. 
The structure coefficients  \eqref{eq: coeff struttura} can be written in terms of the spin connection, considering both Eqs \eqref{eq: tetrad postulate} and \eqref{eq: omega inerziale}:
\begin{equation}\label{eq: f}
    f^{c}{}_{ab} = - (\omega^{c}{}_{ab}-\omega^{c}{}_{ba}).
\end{equation}
Therefore, the definition $\omega^{c}{}_{ab}$ becomes:
\begin{equation}
    \omega^{c}{}_{ab} = \frac{1}{2} (f_{a}^{\text{ c }}{}_{b} + f_{b}^{\text{ c }}{}_a + f^{c}{}_{ab}). 
\end{equation}

It is worth remarking that, in Teleparallel Gravity, one has always to
consider tetrads together with their related spin connections, namely the pair
\(\left\{e^a{}_{\mu},\omega^a{}_{b\mu}\right\}\).

In particular, for proper tetrads the inertial spin connection vanishes,
since they do not carry inertial effects. In this case, the teleparallel pair
reduces to \(\left\{e^a{}_{\mu},0\right\}\)
\cite{Hehl:1994ue,Krssak:2018ywd}. This gauge choice for the spin connection
is the so-called Weitzenb\"ock gauge.
 In this gauge, the tetrad postulate
\(\nabla_{\mu}e^{a}{}_{\nu}=0\) becomes
\begin{equation}\label{eq: weitzenbock}
	\partial_{\mu}e^{a}_{\nu}
	-\Gamma^{\lambda}{}_{\nu\mu}e^{a}_{\lambda}=0.
\end{equation}
This is the \textit{distant parallelism condition}, from where Teleparallel Gravity takes its name. In this gauge, the corresponding Weitzenb\"ock connection and torsion can be written as follows{}
\begin{align}
    \Gamma^{\alpha}_{\phantom{\alpha}\mu\nu}&=e_{a}^{\phantom{a}\alpha}\partial_{\mu}e^{a}_{\phantom{a}\nu},\label{eq: A16} \\
    T^{\alpha}_{\phantom{\alpha}\mu\nu}&=e_{a}^{\phantom{a}\alpha}T^{a}_{\phantom{a}\mu\nu}=e_{a}^{\phantom{a}\alpha}[\partial_{\mu}e^{a}_{\phantom{a}\nu}-\partial_{\nu}e^{a}_{\phantom{a}\mu}] \label{eq: tttn}.
\end{align}

It is particularly interesting to note from the comparison between Eq.$\thinspace$\eqref{eq: t} and Eq.$\thinspace$\eqref{eq: tttn} that the transformation of the vanishing connection (leading to the purely inertial affine connection \eqref{eq: transf coordinate CG}) generates the vierbein in the Weitzenb\"{o}ck gauge \cite{BeltranJimenez:2019esp}.

\subsection{Metric Teleparallel Equivalent of GR}
\label{sec: TEGR}

In Sec.~\ref{Sec: STEGR}, we introduced the framework of symmetric teleparallel theories. We now turn to the metric teleparallel case, characterized by a flat, metric-compatible affine connection. In addition to the flatness condition $R^\alpha{}_{\beta\mu\nu}=0$, given in Eq.~\eqref{eq: flat_curvature}, we therefore impose metric compatibility:
\begin{equation}\label{eq:non-metrpost}
Q_{\alpha\mu\nu}=0.
\end{equation}
In this framework, the independent geometrical object encoding the
gravitational interaction is then the torsion tensor.

The most general even-parity quadratic form constructed from torsion can be
written as the three-parameter combination
\begin{equation}\label{eq: general torsion}
     T_{\rm gen}:=-\frac{c_{1}}{4} T_{\alpha \mu \nu} T^{\alpha \mu \nu}
     -\frac{c_{2}}{2} T_{\alpha \mu \nu} T^{\mu \alpha \nu}
     +c_{3} T_{\alpha} T^{\alpha},
\end{equation}
where \(c_{1},c_{2},c_{3}\) are free real constants, whose explicit values
characterize the so-called gravity model \emph{three-parameter
Hayashi-Shirafuji theory} \cite{Hayashi:1979qx,Blixt:2018znp}. In particular, TEGR is
obtained for \(c_1=c_2=c_3=1\), which gives the torsion scalar appearing in
Eq.\ \eqref{eq:general_torsion},
\begin{align}\label{eq: T tegr}
    T:= \frac{1}{2} S_{a}^{\ \mu\nu}T^{a}_{\ \mu\nu}
    =\frac{1}{4}T^\rho_{\ \ \mu\nu}T_\rho^{\ \mu\nu}+\frac{1}{2}T^\rho_{\ \ \mu\nu}T^{\nu\mu}{}_\rho-T_\mu T^\mu,    
\end{align}
where the tensor \(S_{a}^{\ \mu\nu}\) is the \emph{superpotential}, defined as
\begin{equation}\label{eq:superpotential}
S_{a}{}^{\mu\nu}
:=
K^{\mu\nu}{}_{a}
-e_{a}{}^{\nu} T^\mu
+e_{a}{}^{\mu}T^\nu .
\end{equation}
The torsion scalar contains all independent quadratic contractions of the
torsion tensor and, in Eq.\ \eqref{eq: T tegr}, the first term resembles that
of the usual Lagrangian of internal gauge theories, whereas the other two stem
from the tetrad soldered character, allowing one to set internal and external
indices at the same level \cite{Krssak:2018ywd}.

With the choice of the torsion scalar in Eq.\ \eqref{eq: T tegr}, the general
metric-affine identity \eqref{eq: general R} applies and, using the flatness
condition \eqref{eq: flat_curvature} together with the metric-compatibility
condition \eqref{eq:non-metrpost}, it reduces to
\begin{equation}\label{eq:GR_TEGR}
    \gr{R}=-T-\hat{B}
    =
    -T-\frac{2}{e}\partial_{\mu}\left(e T^{\mu}\right).
\end{equation}

As a consequence of the identity \eqref{eq:GR_TEGR}, the TEGR gravitational action can be defined as
\begin{equation}\label{eq:TEGR_lagrangian}
    S_{\rm TEGR}
    =
    -\frac{c^4}{16\pi G}
    \int_{\mathcal M} {\rm d}^{4}x\ eT ,
\end{equation}
and it differs from the
EH action only by a boundary term. Therefore, it reproduces the same bulk
dynamics as GR. As in the STEGR case, the coupling to matter can be performed
following the usual minimal coupling prescription. For the scalar field
considered below, this implies \(\nabla_\mu\phi=\partial_\mu\phi\). In the
presence of matter, the action \eqref{eq:TEGR_lagrangian} becomes
\begin{equation}\label{eq: TEGR action}
   S_{\rm TEGR}
   =
   -\frac{c^4}{16\pi G}
   \int_{\mathcal M} {\rm d}^{4}x\ eT
   +
   \int_{\mathcal M} {\rm d}^{4}x\ e\,\mathcal{L}_{\rm m}.
\end{equation}
An important issue concerns the coupling to matter fields. In the presence of
torsion, fermionic and bosonic matter couplings can be sensitive to the
distortion of the connection and to the choice of teleparallel frame. These
subtleties can be handled within the covariant formulation of teleparallel
gravity, or equivalently by choosing a suitable Weitzenb\"ock gauge; see
Refs.\ \cite{Obukhov:2004hv,Aldrovandi:2013wha,BeltranJimenez:2019esp,Capozziello:2022zzh}
for more details. In the present work, however, we will only consider a scalar
field minimally coupled to the metric, so that these issues do not affect the
vacuum decay calculation that we will perform in Sec.\,\ref{vacdecayTEGR}.

The metric field equations are obtained by varying the action with respect to
\(g_{\mu\nu}\):
\begin{align}\label{geomT}
\left(\gr{\nabla}_\alpha+T_\alpha \right) S_{(\mu\nu)}{}^\alpha
+t_{\mu\nu}
-\frac 12 Tg_{\mu\nu}
=
\kappa \mathfrak{T}_{\mu\nu}\,,
\end{align}
where
\begin{equation}
 t_{\mu\nu}
 :=
 \frac{\partial T}{\partial g^{\mu\nu}}
 =
 \frac 12 S_{(\mu|}{}^{\lambda \rho} T_{\nu)\lambda \rho}
 -
 T^{\lambda \rho}{}_{(\mu}S_{\lambda \rho|\nu)} ,
\end{equation}
and \(\mathfrak{T}_{\mu\nu}\) is the stress-energy tensor.
Using the Einstein tensor $\gr{G}_{\mu\nu}$,
Eq.\ \eqref{geomT} can be rewritten in the explicit form
\cite{Capozziello:2021pcg}
\begin{align} \label{eq:TEGR_FE_final}
\gr{G}_{\mu\nu}
&:=
\frac{1}{e}e^a{}_{\mu} g_{\nu\rho}
\partial_\sigma(e S_a{}^{\rho\sigma})
-
S_b{}^{\sigma}{}_{\nu}T^b{}_{\sigma\mu}
\notag\\
&\quad
+\frac{1}{2}Tg_{\mu\nu}
-
e^a{}_{\mu}\omega^b{}_{a\sigma}S_{b\nu}{}^\sigma
=
\kappa\mathfrak{T}_{\mu\nu},
\end{align}

The connection field equations, obtained by varying with respect to
\(\Gamma^\alpha{}_{\mu\nu}\), are
\begin{align}\label{connectioneqmT}
\left(\gr{\nabla}_{\alpha}+T_\alpha\right)
\left(\sqrt{-g} S_{[\mu}{}^{\alpha}{}_{\nu]}\right)
=
0\,.
\end{align}
They are identically satisfied as a consequence of the Bianchi identity once
Eq.\ \eqref{geomT} holds \cite{BeltranJimenez:2018vdo}. Therefore, the
connection field equations carry no additional dynamical information, and the
dynamics of TEGR is entirely determined by the Einstein field equations.
Clearly, the torsional boundary term in
Eq.\ \eqref{eq:GR_TEGR} is irrelevant for the derivation of the bulk field
equations under fixed boundary data. As already observed in Sec.~\ref{Sec: STEGR} for the STEGR case, this does not necessarily imply that it
vanishes when evaluated on a specific solution, an issue crucial for the calculation of the transition rate $\Gamma$ of the false vacuum, that is the main objective of the present paper.

Let us finally recall how the metric teleparallel connection is represented.
Since metric compatibility holds, substituting Eq.\ \eqref{eq: gl4} into the
definition of the non-metricity tensor, Eq.\ \eqref{eq: non-metricity}, gives
\begin{equation}\label{eq: inertialTEGRconn}
	\partial_{\alpha}g_{\mu\nu}
	=
	(\Lambda^{-1})^{\rho}_{\phantom{\rho}\lambda}
	\left[
	g_{\rho\nu}\partial_{\alpha}\Lambda^{\lambda}_{\phantom{\lambda}\mu}
	+
	g_{\rho\mu}\partial_{\alpha}\Lambda^{\lambda}_{\phantom{\lambda}\nu}
	\right].
\end{equation}
Equation \eqref{eq: inertialTEGRconn} relates the metric tensor to the affine
connection, giving a description in terms of the 16-component matrix
\(\Lambda\) rather than the 10 independent components of \(g_{\mu\nu}\). This
redundancy is associated with local Lorentz transformations, under which the
theory is invariant
\cite{Heisenberg:2023lru,Capozziello:2023vne,Hohmann:2022mlc}. The torsion
tensor can then be written as
\begin{equation}\label{eq: t}
	T^{\alpha}_{\phantom{\alpha}\mu\nu}
	=
	(\Lambda^{-1})^{\alpha}_{\phantom{\alpha}\lambda}
	\partial_{\mu}\Lambda^{\lambda}_{\phantom{\lambda}\nu}
	-
	(\Lambda^{-1})^{\alpha}_{\phantom{\alpha}\lambda}
	\partial_{\nu}\Lambda^{\lambda}_{\phantom{\lambda}\mu},
\end{equation}
where the relation \eqref{eq: transf coordinate CG} has been used.

After this short review, and in order to pave the way for the calculation of
the false vacuum tunneling rate in TEGR and STEGR, in the next section we
recall the main steps of the vacuum stability analysis within the GR
framework.

\section{Vacuum stability in General Relativity} 
\label{sec:thback}

This section introduces the framework needed to analyze false vacuum decay in STEGR and TEGR at leading semiclassical order. We consider the decay of a Minkowski false vacuum, with vanishing energy density, into an anti-de Sitter (AdS) true vacuum, with negative energy density. We review the corresponding calculation in GR \cite{cdl}, beginning with the flat spacetime case \cite{coleman} in Sec.\ \ref{sec:Minkowski spacetime} and then including gravitational backreaction in Sec.\ \ref{sec:bouncecdl}. In Sec.\ \ref{asymptotic}, we analyze the endpoint behavior of the bounce, which is needed to evaluate the boundary contributions. Finally, in Sections \ref{sec:tunnelingGR} and \ref{sec:tunnelingGR}, we calculate the tunneling exponent $B$, accounting for both the bulk action $S^{\rm EH}$, containing the Einstein-Hilbert and matter terms, and the Gibbons-Hawking-York boundary term $S^{\rm GHY}$.

\subsection{False vacuum decay in flat spacetime background}\label{sec:Minkowski spacetime}
\begin{figure}[t!]
\centering
\includegraphics[width=0.6\textwidth]{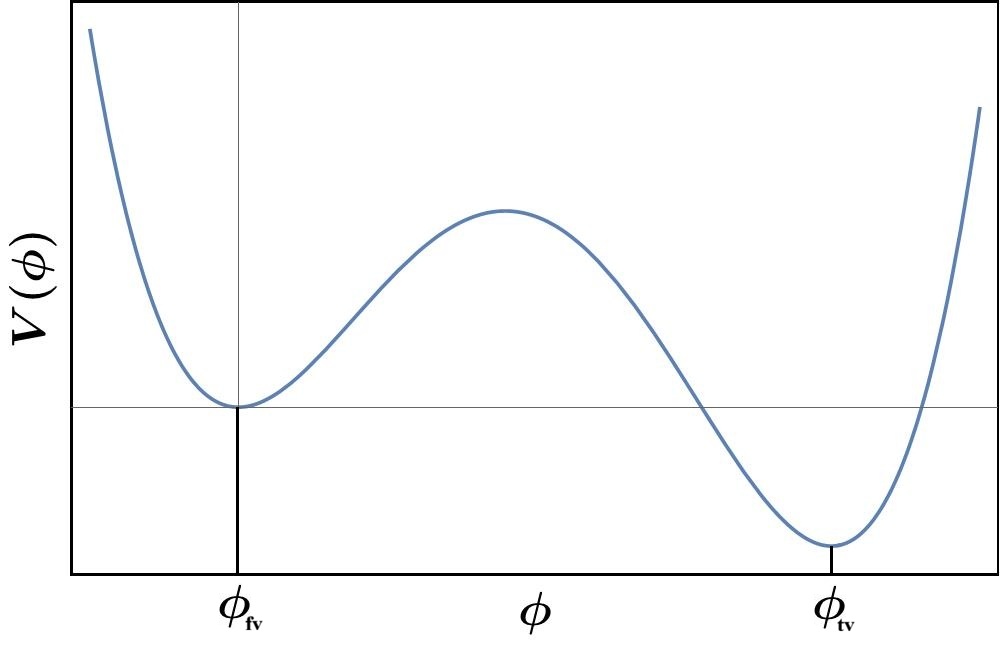}
\caption{
Sketch of the potential $V(\phi)$ described in the text. For $\phi=\phi_{\rm fv}=0$ we have the Minkowski {\it false vacuum}, that corresponds to a local minimum for which $V(\phi_{\rm fv})=0$, while for $\phi=\phi_{\rm tv}$ we have the AdS {\it true vacuum}, that corresponds to the global minimum of the potential.}
\label{fig:potential}
\end{figure}

Let us begin by considering the 
Euclidean action for a single component real scalar field $\phi$ in the Minkowski spacetime background:
\begin{equation}\label{maction}
S[\phi]=\int_{\mathbb R^4} d^4x \left [ \frac 1 2 (\partial_\mu \phi)^2 
+ V(\phi) \right ]\,,
\end{equation}
where $V(\phi)$ is a potential with a local minimum 
(\emph{false vacuum}) at $\phi=\phi_{\rm fv}$, and 
an absolute minimum (\emph{true vacuum}) at $\phi=\phi_{\rm tv}$, see Fig.\,\ref{fig:potential}.  

To calculate the false vacuum decay rate $\Gamma$, we need the Euclidean persistence amplitude of the false vacuum
\begin{equation}
	Z_{\rm fv}
	=
	\int\mathcal D\phi\,e^{-S_E[\phi]},
	\label{eq:flat-path-integral}
\end{equation}
where the integration domain is defined by the condition $\phi(x)\longrightarrow\phi_{\rm fv}$ at the Euclidean infinity. Indicating with $\Omega$ the $4$-volume\, $\Omega=V \,T$, where $V$ is the spatial $3$-volume and $T$ the one-dimensional Euclidean time volume, the decay rate $\Gamma$ is given in terms of $Z_{\rm fv}$ as
\begin{equation}
		\frac{\Gamma}{V}
		=
		\lim_{\Omega\to\infty}
		\frac{2}{\Omega}\,
		\operatorname{Im}\ln Z_{\rm fv}.
	\label{eq:flat-rate-from-Z}
\end{equation}
The persistence amplitude $Z_{\rm fv}$ in \eqref{eq:flat-path-integral} is usually computed in the framework of the semiclassical expansion.
We seek a saddle point of the Euclidean action, satisfying the Euclidean equation of motion $\delta S/\delta\phi=0$, or equivalently,
\begin{equation}\label{eq:phieq}
\nabla_\mu \nabla^\mu \phi=\frac{dV}{d\phi},
\end{equation}
together with the appropriate boundary conditions, specified below. The relevant non-trivial solution, known as the \emph{bounce solution}, is an $O(4)$-symmetric configuration with a radial profile $\phi(x)=\phi_b(r)$, where $r$ denotes the Euclidean radial coordinate \cite{coleman}.

For an $O(4)$-symmetric field configuration, the action \eqref{maction} reduces to
\begin{equation}\label{action}
	S[\phi]
	=
	2 \pi^2 \int_0^\infty d r \ r^3
	\left[
	\frac 1 2 \dot \phi^2 + V(\phi)
	\right],
\end{equation}
where an overdot denotes differentiation with respect to the Euclidean radial coordinate $r$, so that $\dot\phi=d\phi/dr$ and $\ddot\phi=d^2\phi/dr^2$. 
The equation of motion \eqref{eq:phieq} then becomes
\begin{equation}\label{meq}
\ddot \phi(r) + \frac{3}{r} \dot \phi (r) = \frac{d V}{d \phi}.
\end{equation}
The boundary conditions can now be expressed in terms of $r$. The field must approach the false vacuum as $r\to\infty$. At the origin, the requirement that the \emph{friction term} $\frac{3}{r}\dot\phi(r)$ remain finite imposes $\dot\phi(0)=0$. 
Choosing the field origin so that $\phi_{\rm fv}=0$, the bounce therefore satisfies the boundary conditions
\begin{equation}\label{mbc}
\phi(\infty)=\phi_{\rm fv}=0
\qquad
\dot\phi(0)=0.
\end{equation}
The constant false vacuum configuration, $\phi(r)=\phi_{\rm fv}$, also satisfies the equation of motion \eqref{meq} and the boundary conditions \eqref{mbc}.

As stressed in the Introduction, the tunneling exponent $B$ entering the false vacuum decay rate $\Gamma$ is not $S[\phi_b]$ alone,
but rather the difference  
\begin{equation}\label{eq:BMinkowskiRegulated}
B
\equiv
S[\phi_b]
-
S[\phi_{\rm fv}]=
2 \pi^2 \int_0^\infty d r \ r^3
\left[
\frac 1 2 \dot \phi_b^2 + V(\phi_b)
\right]
-
2 \pi^2 \int_0^\infty d r \ r^3 V(\phi_{\rm fv}) \,.
\end{equation}
The complete expression for the decay rate $\Gamma$ of the false vacuum is
\begin{equation}\label{gamma}
\Gamma = D e^{-B},
\end{equation}
where the exponential factor \(e^{-B}\) gives the leading semiclassical contribution, while \(D\) denotes the quantum fluctuation determinant.
For a Minkowski false vacuum, \(V(\phi_{\rm fv})=0\). The false vacuum action
vanishes and the tunneling exponent reduces to $B=S[\phi_b]$.

\subsection{False vacuum decay in curved spacetime background}\label{sec:bouncecdl}

As previously stressed, we are interested in the calculation of the false vacuum
decay rate in the presence of gravity. Here we are considering GR. Starting from \eqref{maction}, the Euclidean action for the scalar field minimally coupled to gravity with the inclusion of the Einstein-Hilbert term reads
\begin{equation}\label{gaction}
S^{^{\rm EH}}[\phi, g_{\mu\nu}]
=
\int_{\mathbb R^4} d^4 x \sqrt{g}
\left[
-\frac{\gr R}{16 \pi G}
+\frac 1 2 g^{\mu \nu}
\gr \nabla_\mu \phi \, \gr \nabla_\nu \phi
+ V(\phi)
\right] .
\end{equation}
For manifolds with boundaries, the action is actually supplemented by
the Gibbons-Hawking-York (GHY) term $S_{\rm GHY}$, so that the total action reads
\begin{align}\label{eq:Stot}
S_{\rm tot}[\phi, g_{\mu\nu}]=S^{^{\rm EH}}[\phi, g_{\mu\nu}]+S^{^{\rm GHY}}[g_{\mu\nu}]\,.
\end{align}
Although the GHY term does not modify the bulk equations of motion, its on-shell value need not vanish.
The tunneling exponent must therefore be calculated from the full action $S_{\rm tot}$, including the GHY contribution.

Let us move now to the bounce solution. As stressed in the previous section, for the theory of a single component scalar field in flat spacetime,
the bounce of minimum action has an $O(4)$ symmetry. The flat-spacetime result does not establish the same conclusion in the presence of gravitational backreaction.  Following the Coleman--De Luccia construction \cite{cdl}, we therefore assume that the dominant bounce retains this symmetry. Given  $O(4)$ invariance, the maximally symmetric Euclidean line element is:
\begin{equation}\label{metric}
ds^2
=
dr^2
+
\rho(r)^2
\left[
d\chi^2
+
\sin^2\chi
\left(
d\theta^2+\sin^2\theta\,d\varphi^2
\right)
\right],
\end{equation}
where \(\rho(r)\) is the radius of the three-sphere at fixed Euclidean radial
coordinate \(r\) \cite{cdl}. The bounce configuration is specified by the
scalar profile \(\phi_b(r)\) and by the metric function \(\rho_b(r)\), which
solve the coupled equations
\begin{subequations}
\begin{align}
&\ddot \phi
+
3\frac{\dot \rho}{\rho}\dot \phi
=
\frac{d V}{d \phi},
\label{gequation}
\\
&\dot \rho^2
=
1
+
\frac{\kappa \rho^2}{3}
\left(
\frac 1 2 \dot \phi^2
-
V(\phi)
\right),
\label{eq:rhodot}
\end{align}
\end{subequations}
where $\kappa\equiv8\pi G$ and overdots denote differentiation with respect to $r$. Equation\ \eqref{gequation} generalizes Eq.\ \eqref{meq} of Sec.\ \ref{sec:Minkowski spacetime}, replacing the flat-spacetime friction coefficient $3/r$ by $3\dot\rho/\rho$. Equation\ \eqref{eq:rhodot} is the $rr$ component of the Einstein equations and provides the gravitational constraint. Similarly to the flat spacetime case, the bounce is specified by a non-trivial solution $(\phi_b(r),\rho_b(r))$ of this system. For the Minkowski false vacuum considered here, the boundary conditions are
\begin{equation}\label{bcond}
\phi_b(\infty)=\phi_{\rm fv},
\qquad
\dot \phi_b(0)=0,
\qquad
\rho_b(0)=0\,,
\end{equation}

As in flat spacetime, the first two conditions impose the approach to the false vacuum at Euclidean infinity and regularity of the scalar field at the origin, respectively.
For the regular asymptotically flat bounce considered here, Eq.\ \eqref{eq:rhodot} implies that $\rho$ vanishes at a finite radial coordinate. Choosing this zero as the radial origin gives $\rho_b(0)=0$, the third condition in Eq.\ \eqref{bcond}. This point defines the center of the bounce, where the three-sphere slices shrink to zero radius. At this regular center, Eq.\ \eqref{eq:rhodot} implies $\dot \rho^2 -  {\ddot \rho} \rho \geq 1$; selecting the branch on which $\rho_b$ increases away from the origin fixes $\dot\rho_b(0)=1$. It is easy to show from this relation that $\rho(r)$ possesses a zero, and for translation invariance we can take $\rho(r=0)=0$. As in the flat spacetime case, we choose the field origin so that $\phi_{\rm fv}=0$.

Asymptotically, \(r\to\infty\), the scalar field approaches the false vacuum
and the geometry approaches the corresponding flat false vacuum geometry. In
particular, the bounce metric function behaves as
\begin{equation}\label{as}
\rho_b(r) \sim r+k_{\rm fv},
\end{equation}
where the constant \(k_{\rm fv}\) is obtained analytically in the thin-wall
regime \cite{cdl}, while in general it is determined by the numerical
integration of Eqs.\ \eqref{gequation} and \eqref{eq:rhodot}. For later
convenience, in the coming subsection we will provide further details on the
asymptotic behavior of the bounce solution.

\subsection{Asymptotic behavior of the bounce}\label{sec:asymptotic-bounce}
\label{asymptotic}

Following \cite{Branchina:2014rva,Bentivegna:2017qry}, we analyze the bounce near $r \to 0$ and as $r \to \infty$. The resulting endpoint behaviors determine the GHY contribution $S_{\rm GHY}$ to $B$ and will also be used in the STEGR and TEGR calculations In Sections \ref{vacdecaySTEGR} and \ref{vacdecayTEGR}, respectively.

Let us start from the equations \eqref{gequation} and \eqref{eq:rhodot}. 
The asymptotic behavior for \(r\to\infty\) is fixed by the fact that the
bounce describes the decay of a Minkowski false vacuum. The false vacuum configuration is
\begin{equation}\label{eq:falsevacuumdef}
\phi=\phi_{\rm fv},
\qquad
\dot\phi=0,
\qquad
V(\phi_{\rm fv})=0 .
\end{equation}
Substituting this configuration into the constraint equation
\eqref{eq:rhodot}, one obtains $\dot\rho_{\rm fv}^{\,2}
=
1 $.
Therefore, on the regular expanding branch we have $\rho_{\rm fv}(r)=r$, 
where the origin has been chosen so that \(\rho_{\rm fv}(0)=0\).
The bounce has to approach this false vacuum solution at large radius. The false vacuum geometry is non-compact, and the
bounce approaches the same non-compact asymptotic branch. Thus, it is $\dot\rho(r)\to1$ and $\rho(r)/r\to1 $.
Equivalently,
\begin{equation}\label{eq:rhoinf}
\rho(r)\sim r
\qquad
(r\to\infty).
\end{equation}
This leading behavior follows from the boundary condition that the bounce
approaches the Minkowski false vacuum solution. The expansion
\(\rho(r)=r+k+\cdots\) will be obtained below from the scalar falloff together
with the constraint equation \eqref{eq:rhodot}.

Let us define
\begin{equation}
\eta(r)\equiv\phi(r)-\phi_{\rm fv},
\qquad
E_\phi(r)\equiv \frac12\dot\phi^2-V(\phi).
\end{equation}
Since the bounce approaches the false vacuum, we have
\begin{equation}
\eta(r)\to0,
\qquad
\dot\phi(r)\to0,
\qquad
E_\phi(r)\to0 .
\end{equation}
Moreover, Eq.\ \eqref{eq:rhoinf} gives
\begin{equation}
H(r)\equiv\frac{\dot\rho}{\rho}
\sim
\frac1r ,
\end{equation}
and the scalar equation \eqref{gequation} becomes
\begin{equation}\label{eq:asymptotic_eta_equation}
\ddot\eta+3H(r)\dot\eta
=
V'(\phi_{\rm fv}+\eta).
\end{equation}
We now distinguish two cases.
If the false vacuum is non-degenerate, namely $m_{\rm fv}^2\equiv V''(\phi_{\rm fv})>0$,
then $V'(\phi_{\rm fv}+\eta)
=
m_{\rm fv}^2\eta
+
O(\eta^2)$.
Keeping the leading asymptotically flat friction term \(H(r)\sim 1/r\),
Eq.\ \eqref{eq:asymptotic_eta_equation} becomes
\begin{equation}
\ddot\eta+\frac{3}{r}\dot\eta
\simeq
m_{\rm fv}^2\eta .
\end{equation}
The exponentially growing solution of the above equation is excluded by the false vacuum boundary
condition, while the decaying solution has the radial
falloff
\begin{equation}\label{eq:phimass}
\eta(r)
\sim
C\,\frac{e^{-m_{\rm fv}r}}{r^{3/2}} .
\end{equation}
Equivalently, for every \(\mu<m_{\rm fv}\), we have $\eta(r)=O(e^{-\mu r})$ and $E_\phi(r)=O(e^{-2\mu r})$.
Using this falloff in the constraint equation \eqref{eq:rhodot}, together with
the leading behavior \(\rho(r)\sim r\), we get
\begin{equation}\label{eq:dotrhomass}
\dot\rho^2-1
=
O(r^2e^{-2\mu r}) \quad \Rightarrow \quad \dot\rho-1
=
O(r^2e^{-2\mu r}),\,
\end{equation}
since \(\dot\rho\to1\).
This falloff is integrable at infinity. Therefore the leading asymptotic
behavior is
\begin{equation}\label{rhoinfmassive}
\rho(r)
=
r+k
+
O(r^2e^{-2\mu r}) .
\end{equation}
In particular,
\begin{equation}\label{eq:weighted_rho_falloff_massive}
\rho^2(\dot\rho-1)
=
O(r^4e^{-2\mu r})
\longrightarrow0 .
\end{equation}

Let us move now to the massless case, $V''(\phi_{\rm fv})=0$.
The first non-vanishing term in the potential is quartic. We then have: $V'(\phi_{\rm fv}+\eta)=O(\eta^3)$.
Under these conditions, the leading asymptotic scalar equation \eqref{eq:asymptotic_eta_equation} becomes
\begin{equation}
\ddot\eta+\frac{3}{r}\dot\eta\simeq0 \,,
\end{equation}
whose solution compatible with \(\eta\to0\) is
\begin{equation}\label{phiinf}
\eta(r)
=
\frac{c_\phi}{r^2}
+
O(r^{-3}) .
\end{equation}
Since the leading term is quartic, $V(\phi)=O(r^{-8})$ and $
E_\phi(r)
=
\frac{2c_\phi^2}{r^6}
+
O(r^{-7})$. Introducing this latter result in Eq.\,\eqref{eq:rhodot}, together with the leading behavior
\(\rho(r)\sim r\), one
finds
\begin{equation}\label{eq:dotrhomassless}
\dot\rho^2-1
=
\frac{2\kappa c_\phi^2}{3r^4}
+
O(r^{-5}),
\quad \Rightarrow \quad
\dot\rho-1
=
\frac{\kappa c_\phi^2}{3r^4}
+
O(r^{-5}) .
\end{equation}
This falloff is integrable at infinity, so that
\begin{equation}\label{aasympt}
\rho(r)
=
r+k
-
\frac{\kappa c_\phi^2}{9r^3}
+
O(r^{-4}) .
\end{equation}
In particular,
\begin{equation}\label{eq:weighted_rho_falloff_massless}
\rho^2(\dot\rho-1)
=
O(r^{-2})
\longrightarrow0 .
\end{equation}

Now we study the behavior for $r\to 0$. Since the boundary conditions \eqref{bcond} impose that the bounce solution is a regular function in $r=0$, we can consider power expansions for $\phi(r)$ and $\rho(r)$
\begin{align}\label{expansion}
\phi(r)=A_0+A_2r^2+A_3r^3+\dots \qquad \rho (r)=B_1r+B_2r^2+B_3r^3+\dots\,,
\end{align}
where the linear term $A_1r$ in $\phi(r)$ and the constant term $B_0$ in $\rho(r)$ are not present due to  the conditions $\dot\phi(0)=0$ and $\rho(0)=0$, respectively. Substituting \eqref{expansion} in Eqs.\,\eqref{gequation} and \eqref{eq:rhodot}, and comparing the coefficients of the powers of $r$, we get a set of algebraic equations for $A_i$ and $B_i$ ($i>0$), which can be written in terms of $A_0$
\begin{align}
B_1=1\,, \qquad B_2&=0\,, \qquad B_3=-\frac{\kappa}{18}V(A_0)\,, \quad \dots \\
A_2=\frac 18 V'(A_0)\,, \qquad A_3&=0\,, \qquad A_4=\frac{1}{576}V'(A_0)\left[ 2\kappa V(A_0)+3V''(A_0) \right]\,, \quad \dots
\end{align}
In conclusion, for the bounce solution around $r=0$ we have
\begin{align}\label{rto0}
\phi(r) =A_0+\frac 18 V'(A_0)\,r^2+\dots \qquad ; \qquad \rho(r)=r-\frac{\kappa}{18}V(A_0)\,r^3+\dots \,.
\end{align}
Up to now, to find the behavior of $\phi(r)$ and $\rho(r)$ in the vicinity of $r=0$, we have used only the boundary conditions \eqref{bcond}$_2$ and \eqref{bcond}$_3$. This is why in \eqref{rto0} we still have an arbitrary constant $A_0$. To fix the latter, we have to use the condition \eqref{bcond}$_1$, i.e. the behavior of $\phi(r)$ at the infinite endpoint. The technique to determine $A_0$ is the well-know Coleman's {\it overshoot-undershoot method} \cite{coleman}: since the bounce is the only solution to the field equations \eqref{gequation} and \eqref{eq:rhodot} 
 that presents the asymptotic behavior ($r \to \infty$) given by Eq.\,\eqref{eq:phimass} (for the massive case) or Eq.\,\eqref{phiinf} (for the massless case), the unique value $\bar A_0$ such that this asymptotic behavior is fulfilled is determined by varying $A_0$.
 
In the next subsection, we present the derivation of the tunneling exponent $B$, considering the contribution $B^{^{\rm EH}}$ coming from the bulk action of the Einstein-Hilbert matter action, $S^{^{\rm EH}}$,  and the contribution $B^{^{\rm GHY}}$ coming from the boundary term of the GHY action, $S^{^{\rm GHY}}$.

\subsection{Calculation of the tunneling exponent \texorpdfstring{$B$}{B}}\label{sec:tunnelingGR}

To calculate the tunneling exponent $B$, we have to consider the total action $S_{\rm tot}[\phi, g_{\mu\nu}]$ in \eqref{eq:Stot}, that consists of the Einstein-Hilbert (plus matter) term $S^{^{\rm EH}}$ and the boundary term $S^{^{\rm GHY}}$, and evaluate it on-shell, at bounce solution and at the false vacuum. 

Let us begin with $S^{^{\rm EH}}$ evaluated at the bounce ($S_b^{^{\rm EH}}$) and at the false vacuum ($S_{\rm fv}^{^{\rm EH}}$), and calculate the contribution $B^{^{\rm EH}}$ to $B$, i.e. 
$B^{^{\rm EH}}=S_b^{^{\rm EH}}-S^{^{\rm EH}}_{\rm fv}$.
Considering the
(Euclidean) stress-energy tensor $	\mathfrak{T}_{\mu \nu}
=
\partial_\mu \phi\, \partial_\nu \phi
-
g_{\mu \nu}
\left(
\frac 1 2 \partial_\sigma \phi\,\partial^\sigma \phi
+
V(\phi)
\right)$, and taking the trace of the Einstein equation
\(\gr G_{\mu\nu}=\kappa\mathfrak{T}_{\mu\nu}\), the Ricci scalar can be written as
\begin{equation}\label{trace}
	\frac 1 \kappa \,\gr R
	=
	g^{\mu\nu}\partial_\mu\phi\,\partial_\nu\phi
	+
	4V(\phi).
\end{equation}
Inserting \eqref{trace} into \eqref{gaction}, and using the $O(4)$ metric~\eqref{metric}, the on-shell action becomes
\begin{equation}\label{gravactioncutoff}
	S^{^{\rm EH}}[\phi,\rho]
	=
	-2\pi^2
	\int_0^{\infty} dr\,\rho^3\,V(\phi),
\end{equation}
so that the tunneling exponent $B^{^{\rm EH}}$ is
\begin{equation}\label{eq:BGRonShell}
	B^{^{\rm EH}}
	=
	-2\pi^2
	\int_0^{\infty} dr\,\rho_b^3\,V(\phi_b)
	+
	2\pi^2
	\int_0^{\infty} dr\,\rho_{\rm fv}^3\,V(\phi_{\rm fv}) .
\end{equation}
Since we are considering the Minkowski false vacuum (\(V(\phi_{\rm fv})=0\) and
\(\rho_{\rm fv}=r\)), the second term in the right hand side of \eqref{eq:BGRonShell} vanishes. 

We now move to the boundary (Gibbons-Hawing-York) term $S^{^{\rm GHY}}$ (below we write this term for a generic manifold $\mathcal M$. In the present work we have $\mathcal M=\mathbb R^4$), 
\begin{equation}
	S^{^{\rm GHY}}
	=-\frac{1}{\kappa}
	\int_{\partial\mathcal M}\dd^3x\,\sqrt h\,K,
	\label{eq:GHY-definition}
\end{equation}
where $\partial\mathcal M$ is the boundary surface of the manifold $\mathcal M$, $h_{ij}$ the induced metric on $\partial\mathcal M$, $h_{ij}=g_{\mu\nu}\frac{\partial x^\mu}{\partial y^i}\frac{\partial x^\nu}{\partial y^j}$ (where $i,j=1,2,3$ and $y^i$ are the intrinsic coordinate on the boundary surface $\partial\mathcal M$), and $K$ the extrinsic curvature, $K=\nabla_\mu n^\mu$,  with $n^\mu$ unitary vector normal to the surface (outgoing). When evaluated on the bounce and on the false vacuum, the corresponding boundary actions $S^{^{\rm GHY}}_b$ and $S^{^{\rm GHY}}_{\rm fv}$  diverge separately in the limit in which the boundary is sent to infinity (see below).
Therefore, in order to compute the contribution $B^{^{\rm GHY}}=S^{^{\rm GHY}}_b-S^{^{\rm GHY}}_{\rm fv}$ to the tunneling exponent $B$,
one has first to regularize both  $S^{^{\rm GHY}}_b$ and $S^{^{\rm GHY}}_{\rm fv}$. In our case $\mathcal M=\mathbb R^4$ ($O(4)$ symmetry), and the natural regularization is to consider as \vv cutoff" boundary a three-sphere $S^3$ of finite radius $\bar \rho$. Clearly, to have a consistent regularization, both  $S^{^{\rm GHY}}_b$ and $S^{^{\rm GHY}}_{\rm fv}$ have to be calculated with the same $\bar \rho$. Since $\rho_b(r)$ and  $\rho_{\rm fv}(r)$ are different functions of the radial coordinate $r$, the \vv cutoff coordinate" for the bounce, say $\bar r_b$, must be different from the \vv cutoff coordinate" for the false vacuum, say $\bar r_{\rm fv}$, so that
\begin{equation}\label{eq:matchedcutoff}
	\rho_b(\bar r_b)=\rho_{\rm fv}(\bar r_{\rm fv})=\bar\rho .
\end{equation}
$B^{^{\rm GHY}}$ is obtained taking the limit for $\bar \rho \to \infty$ of the regularized difference $(S^{^{\rm GHY}}_b)_{\rm reg}-(S^{^{\rm GHY}}_{\rm fv})_{\rm reg}$.

 Let us begin with $S^{^{\rm GHY}}_b$, the boundary action evaluated at the bounce. In this case, the extrinsic curvature is $K
=\frac{1}{\sqrt g}\partial_r\sqrt g
\notag=\frac{3\dot \rho}{\rho}$ and, since
	$\int_{S^3}\dd^3x\,\sqrt h=2\pi^2 \bar \rho^{\,3}$,
we have
\begin{align}
(S^{^{\rm GHY}}_b)_{\rm reg}=	S^{^{\rm GHY}}[\rho_b;\bar r_b]
	=-\frac{1}{\kappa}
	\int_{S^3}\dd^3x\,\sqrt h\,K
	=-\frac{6\pi^2}{\kappa}
	 \rho_b^2(\bar r_b)\dot \rho_b(\bar r_b)=-\frac{6\pi^2}{\kappa}
	 \bar \rho^2\dot \rho_b(\bar r_b).
	\label{eq:bounce-GH}
\end{align}
Going to the false vacuum solution, $K_{\rm fv}=\frac{3}{\bar \rho}=\frac{3}{\rho_{\rm fv}(\bar r_{\rm fv})}$ and then
\begin{equation}
		(S^{^{\rm GHY}}_{\rm fv})_{\rm reg}=S^{^{\rm GHY}}[\rho_{\rm fv};\bar r_{\rm fv}]
		=-\frac{6\pi^2}{\kappa}\rho^2_{\rm fv}(\bar r_{\rm fv})=-\frac{6\pi^2}{\kappa}\bar \rho^2.
	\label{eq:false-GHY}
\end{equation}
The massive case in Eqs.\,\eqref{rhoinfmassive}  and massless one \eqref{aasympt} show that $\bar r_b \to \infty$ when $\bar \rho \to \infty$. Therefore, Eq.\ \eqref{eq:bounce-GH}, together with the massive and massless cases of Eqs.\ \eqref{eq:dotrhomass} and \eqref{eq:dotrhomassless}, respectively, show that $(S^{^{\rm GHY}}_{b})_{\rm reg}$ diverges for $\bar \rho \to \infty$. The same holds true also for $(S^{^{\rm GHY}}_{\rm fv})_{\rm reg}$, because of Eq.\ \eqref{eq:false-GHY} .

Subtracting \eqref{eq:false-GHY} to \eqref{eq:bounce-GH} and using \eqref{eq:matchedcutoff}, we have
\begin{align}
	(S^{^{\rm GHY}}_{\rm b})_{\rm reg}
	-
	(S^{^{\rm GHY}}_{\rm fv})_{\rm reg}
	=
	-\frac{6\pi^2}{\kappa}
	\bar\rho^2
	\left[
	\dot\rho_b(\bar r_b)-1
	\right]=
	-\frac{6\pi^2}{\kappa}
	\rho_b^2(\bar r_b)
	\left[
	\dot\rho_b(\bar r_b)-1
	\right]\,.
	\label{eq:GHY-difference}
\end{align}
As stressed above $\bar r_b \to \infty$ when $\bar \rho \to \infty$, and the contribution $B^{^{\rm GHY}}$ to $B$ is then given by
\begin{equation}\label{eq:BGHYzero}
	B^{^{\rm GHY}}
	=
	\lim_{\bar\rho\to\infty}
	\left[ (S^{^{\rm GHY}}_{\rm b})_{\rm reg}
	-
	(S^{^{\rm GHY}}_{\rm fv})_{\rm reg} \right]=
	\lim_{\bar\rho\to\infty} \left[ -\frac{6\pi^2}{\kappa}
	\rho_b^2(\bar r_b)
	\left[
	\dot\rho_b(\bar r_b)-1
	\right] \right]=0\,.
\end{equation}
The asymptotic falloffs in Eqs.\ \eqref{eq:weighted_rho_falloff_massive} and \eqref{eq:weighted_rho_falloff_massless} establish that $B^{\rm GHY}=0$ for the Minkowski-to-AdS decay considered here. In Sections\ \ref{vacdecaySTEGR} and \ref{vacdecayTEGR}, we use the same endpoint analysis to calculate the tunneling exponents in STEGR and TEGR.

\section{Vacuum stability in STEGR}
\label{vacdecaySTEGR} 

Our goal now is to study the false vacuum decay when gravity is described by STEGR. Let us begin by considering the STEGR action \eqref{eq:STEGR_Lagrangian} in the
Euclidean setting. Denoting with
\(Q_E\) the Euclidean version of the non-metricity scalar
\eqref{eq: non-metricity scalar}, the Euclidean action for the scalar field minimally coupled to gravity, including the pure gravity $-\frac{Q_E}{2\kappa}$ term, reads
\begin{equation}\label{STEGRbounceaction}
S_E[\phi,g,\Gamma]
=
\int_{\mathbb R^4}d^4x\sqrt{g_E}
\left[
-\frac{Q_E}{2\kappa}
+\frac{1}{2}g_E^{\mu\nu}\nabla_\mu\phi\nabla_\nu\phi
+V(\phi)
\right].
\end{equation}
Since from now on we work only in Euclidean signature, we drop the subscript
\(E\).

The geometric identity
\eqref{eq:GR_STEGR} continues to hold in the Euclidean theory, since it is a
purely geometric relation involving the symmetric affine connection
$\Gamma^\alpha{}_{\mu\nu}
=
\gr{\Gamma}^\alpha{}_{\mu\nu}
+
L^\alpha{}_{\mu\nu}$.
Moreover, since \(\phi\) is a scalar field, its covariant derivative reduces
to the ordinary partial derivative,
$\nabla_\mu\phi=\partial_\mu\phi$,
so the matter sector does not distinguish between the Levi-Civita and
the symmetric teleparallel connections.
	Therefore, the Euclidean STEGR action is dynamically equivalent to the
	Euclidean GR action \eqref{gaction}, in the same sense as in Lorentzian
	signature \cite{BeltranJimenez:2020sih,Oshita:2017nhn}. Due to
	Eq.\ \eqref{eq:GR_STEGR}, the two formulations differ by a total derivative,
	so that
	\begin{equation}\label{eq:VV}
		S_{_{\rm STEGR}}[\phi,g,\Gamma]
		=
		S_{_{\rm GR}}[\phi,g]
		+
		\int_{\mathbb R^4} d^4x\,\partial_\mu V^\mu ,
	\end{equation}
	where \(V^\mu=V^\mu(g,\Gamma)\) is the non-metricity current that can be read
	from the second term on the right hand side of Eq.\ \eqref{eq:GR_STEGR}. Clearly,
	the second term in Eq.\ \eqref{eq:VV} does not modify the field equations in
	the interior of the Euclidean manifold.
	
	For the vacuum stability problem, this bulk equivalence implies that STEGR
	and GR possess the same \(O(4)\)-symmetric bounce solution (seen in the previous section),
	\begin{equation}
		\{\phi_b(r),\rho_b(r)\}_{_{\rm STEGR}}
		=
		\{\phi_b(r),\rho_b(r)\}_{_{\rm GR}},
	\end{equation}
	satisfying the same equations of motion, Eqs.\ \eqref{gequation} and
	\eqref{eq:rhodot}, and the same boundary conditions \eqref{bcond}.
	
	However, equality of the bulk saddle does not necessarily imply equality of
	the corresponding on-shell Euclidean actions. This point is particularly
	relevant for the vacuum decay problem considered here. The bounce describing
	the decay of a Minkowski false vacuum is non-compact, and the tunneling
	exponent \(B\) is defined by the difference between the Euclidean action of
	the bounce and that of the false vacuum reference configuration. Therefore,
	in order to determine the role played by boundary term in Eq.\ \eqref{eq:VV}, one has to
	evaluate at the bounce the corresponding surface contribution on a large but finite
	three-sphere, match this boundary with the false vacuum one, and then study
	the asymptotic limit in which the boundary is sent to infinity. The relevant step is the asymptotic analysis of the bounce, more precisely the determination of
	its asymptotic falloff. The realization of this analysis is one of the main goals of the present work.
	Only after that one can establish whether the
	boundary term contributes or not to the tunneling exponent. 

To proceed with the determination of  $B$, we first need to calculate the most general flat and torsionless affine connection compatible with the \(O(4)\) symmetry of the bounce
\cite{Hohmann:2019fvf,Hohmann:2024phz}. This is what we do in the next section.
  
\subsection{\texorpdfstring{$\boldsymbol{O(4)}$}{O(4)}-invariant affine connection}
\label{affineconnectioncalc}
We now derive the most general $O(4)$-invariant affine connection, exploiting the symmetries of the problem: the invariance of $\Gamma^\alpha{}_{\mu\nu}$ along the flux of the Killing vectors for the $SO(4)$ symmetry, and invariance under reflection transformations\footnote{We will follow the procedure outlined in \cite{Hohmann:2019nat,Hohmann:2019fvf}, where the same calculations were performed to find the most general $O(3)$ invariant affine connection in Lorentzian manifolds.}. Then, since we are working in STEGR, we have also to impose the flat and torsionless postulates, Eqs.\ \eqref{eq: flat_curvature} and \eqref{eq:torspost} respectively.
Finally, having the linear affine connection, we will be able to determine the tunneling exponent, i.e.\,\,the action \eqref{STEGRbounceaction} in the bounce solution.

In particular, due to the $O(4)$ symmetry, it is convenient to use the spherical coordinate system $(r,\chi,\theta,\phi)$ (see for instance the metric \eqref{metric}). At this point, it is important to highlight that,
although working in coincident gauge has no impact on the field equations, we cannot use it, i.e. $\Gamma^\alpha{}_{\mu\nu}=0$, since it is not generally $O(4)$ invariant.

The $SO(4)$ symmetry of the problem is exploited looking for the Lie derivatives of the metric $g_{\mu\nu}$ and the linear affine connection $\Gamma^\alpha{}_{\mu\nu}$, along the  Killing vectors $X^\mu$ (see Eq.\,\eqref{Killing}). In particular, for the metric \eqref{metric} we get
\begin{align}
\mathcal{L}_X g_{\mu\nu} = X^\lambda \partial_\lambda g_{\mu\nu} + g_{\lambda\nu} \partial_\mu X^\lambda + g_{\mu\lambda} \partial_\nu X^\lambda=0\,,
\end{align}
for each of the Killing vectors  $X^\mu$. Moreover, considering all the possible reflection transformations (see Eqs.\,\eqref{refl1}-\eqref{refl4}),
we see that Eq.\,\eqref{metric} is actually the most general $O(4)$-invariant metric.

In order to find the most general $SO(4)$ invariant linear connection, we have to solve the equations
\begin{align}\label{GammaLie}
(\mathcal{L}_X \Gamma)^\alpha_{\mu\nu} = X^\sigma \partial_\sigma \Gamma^\alpha_{\mu\nu} - \Gamma^\sigma_{\mu\nu} \partial_\sigma X^\alpha + \Gamma^\alpha_{\sigma\nu} \partial_\mu X^\sigma + \Gamma^\alpha_{\mu\sigma} \partial_\nu X^\sigma + \partial_\mu \partial_\nu X^\alpha=0, \quad \quad \forall X^\mu.
\end{align}
First of all, the Lie derivative
\begin{align}
(\mathcal{L}_{X_1} \Gamma)^\alpha_{\mu\nu} = -\partial_\phi \Gamma^\alpha_{\mu\nu}=0
\end{align}
implies that all the connection components are $\phi$-independent. Moreover, instead of solving the equations $(\mathcal{L}_{X_2} \Gamma)^\alpha_{\mu\nu}=0$ and $(\mathcal{L}_{X_4} \Gamma)^\alpha_{\mu\nu}=0$, we can consider the linear combinations
\begin{align}\label{lin1}
\sin \phi \,(\mathcal{L}_{X_2} \Gamma)^\alpha_{\mu\nu}-\cos \phi \, (\mathcal{L}_{X_4} \Gamma)^\alpha_{\mu\nu}&=0\,, \\
\label{lin2}
\cos \phi \,(\mathcal{L}_{X_2} \Gamma)^\alpha_{\mu\nu}+\sin \phi \, (\mathcal{L}_{X_4} \Gamma)^\alpha_{\mu\nu}&=0 \,.
\end{align}
Equations \eqref{lin1} give a set of $44$ algebraic equations: with straightforward but tedious calculations, $29$ connection components turn out to be zero, while $15$ are expressed as combinations of trigonometric functions of the $\theta$ coordinate and of the remaining $20$ independent components. Instead, from Eq. \eqref{lin2} we get a set of $16$ differential equations, whose solution fully fix the $\theta$-dependence of the connection components. From the Lie derivative
\begin{align}
(\mathcal{L}_{X_6} \Gamma)^\alpha_{\mu\nu}=0\,,
\end{align}
we get a set of $20$ algebraic and differential equations that fully fix the $\chi$-dependence, and leave the connection components in terms of five arbitrary functions $c_i(r)$ ($i=1, \dots 5$).
Finally, the equations $(\mathcal{L}_{X_3} \Gamma)^\alpha_{\mu\nu}=0$ and $(\mathcal{L}_{X_5} \Gamma)^\alpha_{\mu\nu}=0$ are identically satisfied.
In order to obtain an $O(4)$-invariant connection, we have to consider the reflection transformations in Eqs.\,\eqref{refl1}-\eqref{refl4}, from which we read the following rules: (i) if the number of coordinate indices $\chi$, $\theta$ or $\phi$ in $\Gamma^\alpha{}_{\mu\nu}$ is odd, a factor $-1$ is incurred, (ii) additional factors $-1$ arises depending on the parity under such transformations of the trigonometric functions of $\chi$ and $\theta$. Thus, it turns out that $c_5(r)=0$ and the most general $O(4)$-invariant connection is
\begin{align}\label{O4}
\Gamma^r{}_{\mu\nu}&=
\begin{pmatrix}
c_1(r) & 0 & 0 & 0 \\
0 & c_2(r) & 0 & 0 \\
0 & 0 & c_2(r)\sin^2\chi & 0 \\
0 & 0 & 0 & c_2(r)\sin^2\chi \sin^2\theta \\
\end{pmatrix}
\,,\nn \\
\Gamma^\chi{}_{\mu\nu}&=
\begin{pmatrix}
0 & c_3(r) & 0 & 0 \\
c_4(r) & 0 & 0 & 0 \\
0 & 0 & -\sin\chi \cos \chi & 0 \\
0 & 0 & 0 & -\sin\chi \cos \chi\sin^2\theta \\
\end{pmatrix}
\,,
\nn  \\
\Gamma^\theta{}_{\mu\nu}&=
\begin{pmatrix}
0 & 0 & c_3(r) & 0 \\
0 & 0 & \cot \chi & 0 \\
c_4(r) & \cot \chi & 0 & 0 \\
0 & 0 & 0 & -\sin\theta \cos\theta \\
\end{pmatrix}
\,, \,\,\,
\Gamma^\phi{}_{\mu\nu}=
\begin{pmatrix}
0 & 0 & 0 & c_3(r) \\
0 & 0 & 0  & \cot \chi \\
0 & 0 & 0 & \cot \theta \\
c_4(r) & \cot \chi & \cot \theta & 0 \\
\end{pmatrix}
\,.
\end{align}

Now we proceed to implement the flat and torsionless postulates, Eqs.\,\eqref{eq: flat_curvature} and \eqref{eq:torspost}.
The torsionless condition translates into a symmetric connection under $\mu \leftrightarrow \nu$, which implies 
$c_4(r)=c_3(r)$. 
Finally, the addition of the teleparallel constraint, i.e. $R^\alpha{}_{\mu\nu\lambda}=0$, to the previous 
$T^{\alpha}{}_{\mu\nu}=0$, infers to the following set of equations:
\begin{align}\label{eq1}
1+c_2(r)c_3(r)&=0 \,,\\
\label{eq2}
\dot c_2(r)+c_2(r)c_1(r)-c_2(r)c_3(r)&=0 \,, \\
\label{eq3}
\dot c_3(r)+c_3^2(r)-c_1(r)c_3(r)&=0 \,.
\end{align}
From Eq.\,\eqref{eq1} we have $c_3(r)=-1/c_2(r)$, so that Eqs.\,\eqref{eq2} and \eqref{eq3} both reduce to
\begin{align}
\dot c_2(r)+c_2(r)c_1(r)+1=0 \qquad \Rightarrow \qquad c_1(r)=-\frac{1+\dot c_2(r)}{c_2(r)}\,.
\end{align}
Thus, denoting $c(r)\equiv c_2(r)$, the components of the most general $O(4)$ invariant, flat, and torsionless connection are

\begin{align}\label{connection}
\Gamma^r{}_{\mu\nu}&=
\begin{pmatrix}
-\frac{1+\dot c(r)}{c(r)} & 0 & 0 & 0 \\
0 & c(r) & 0 & 0 \\
0 & 0 & c(r)\sin^2\chi & 0 \\
0 & 0 & 0 & c(r)\sin^2\chi \sin^2\theta \\
\end{pmatrix}
\,,\nn \\
\Gamma^\chi{}_{\mu\nu}&=
\begin{pmatrix}
0 & -\frac{1}{c(r)} & 0 & 0 \\
-\frac{1}{c(r)} & 0 & 0 & 0 \\
0 & 0 & -\sin\chi \cos \chi & 0 \\
0 & 0 & 0 & -\sin\chi \cos \chi\sin^2\theta \\
\end{pmatrix}
\,,
\nn
\\
\Gamma^\theta{}_{\mu\nu}&=
\begin{pmatrix}
0 & 0 & -\frac{1}{c(r)} & 0 \\
0 & 0 & \cot \chi & 0 \\
-\frac{1}{c(r)} & \cot \chi & 0 & 0 \\
0 & 0 & 0 & -\sin\theta \cos\theta \\
\end{pmatrix}
\,, \,\,\,
\Gamma^\phi{}_{\mu\nu}=
\begin{pmatrix}
0 & 0 & 0 & -\frac{1}{c(r)} \\
0 & 0 & 0 & \cot \chi \\
0 & 0 & 0 & \cot \theta \\
-\frac{1}{c(r)} & \cot \chi & \cot \theta & 0 \\
\end{pmatrix}
\,.
\end{align}
Finally, we can calculate the non-metricity scalar, Eq. \eqref{eq: non-metricity},  using the superpotential $P^{\alpha}{}_{\mu\nu}$ of Eq. \eqref{eq:superpotSTEGR} and the connection components of Eq. \eqref{connection}:
\begin{equation}\label{QO(4)}
Q=\frac{3 \left(2 \dot \rho(r)^2+\dot c(r)+2\right)}{\rho(r)^2}+\frac{3 c(r) \dot \rho(r)}{\rho(r)^3}+\frac{9 \dot \rho(r)}{\rho(r) c(r)}-\frac{3 \dot c(r)}{c(r)^2}\,;
\end{equation}
the two independent traces of the non-metricity tensor, $Q_{\alpha}$ and $\tilde{Q}_{\alpha}$ given in Eqs. \eqref{eq:Qa} and \eqref{eq:tildeQa}, have only one non-zero component:
\begin{align}\label{QQtraces}
Q_r=\frac{6 \dot \rho(r)}{\rho(r)}+\frac{2 \left(\dot c(r)+4\right)}{c(r)} \qquad \tilde Q_r=\frac{5 \rho(r)^2+2\rho(r)^2 \dot c(r)-3 c(r)^2}{c(r)\rho(r)^2}\,.
\end{align}

The expressions above show that the Euclidean STEGR connection compatible with
the \(O(4)\) symmetry is not fixed uniquely by the metric. After imposing
flatness and vanishing torsion, a residual function \(c(r)\) remains. This
function is not a new dynamical degree of freedom in the bulk, but a residual
affine data \cite{Blixt:2023kyr,Ferrara:2026hbq}. Nevertheless, since the STEGR action differs from the GR one by a
boundary term, its behavior at the endpoints of the radial geometry has to be
specified before evaluating the on-shell action.

The origin of this residual freedom is simple. Since the STEGR connection is
flat and torsionless, it is locally pure gauge. In the \(O(4)\)-symmetric
sector one may introduce affine coordinates of the form
\begin{equation}\label{eq:affine_coordinates_s}
\xi^a=s(r)n^a(\chi,\theta,\varphi),
\end{equation}
where \(n^a\) is a unit radial vector on the three-sphere $S^3$. Comparing the pure-gauge connection
generated by \eqref{eq:affine_coordinates_s} with the general connection
\eqref{connection}, one finds
\begin{equation}\label{eq:c_from_affine_map}
c(r)
=
-\frac{s(r)}{\dot s(r)} .
\end{equation}
Thus \(c(r)\) parametrizes the remaining freedom in the choice of the affine
radial representative.

Let us first consider the regular endpoint \(r=0\). Regularity and local
invertibility of the affine map require
\begin{equation}\label{eq:affine_origin_regular}
s(0)=0,
\qquad
\dot s(0)\neq0 .
\end{equation}
Writing
\begin{equation}\label{eq:s_origin_expansion}
s(r)=a r+b r^2+O(r^3),
\qquad
a\neq0,
\end{equation}
Eq.\ \eqref{eq:c_from_affine_map} gives
\begin{equation}\label{eq:c_origin_expansion}
c(r)
=
-r
+
\frac{b}{a}r^2
+
O(r^3).
\end{equation}
Therefore,
\begin{equation}
c(r)=-r+O(r^2),
\qquad
\dot c(r)=-1+O(r),
\qquad
(r\to0).
\end{equation}
This behavior follows from regularity of the affine gauge map, not from
regularity of the metric alone.

We now turn to the asymptotic endpoint. Since the tunneling exponent is
defined by comparing the bounce with the false vacuum reference
configuration, the two configurations have to be evaluated with the same
boundary prescription. For the non-compact geometry considered here, this is
implemented by introducing a large physical three-sphere of radius
\(\bar\rho\) and by choosing two radial positions, \(\bar r_b\) and
\(\bar r_{\rm fv}\), such that
\begin{equation}\label{eq:boundary}
	\rho_b(\bar r_b)
	=
	\rho_{\rm fv}(\bar r_{\rm fv})
	=
	\bar\rho .
\end{equation}
In other words, the comparison is performed at the same physical boundary,
not necessarily at the same value of the radial coordinate.

In STEGR, the same idea has to be applied also to the asymptotic affine
sector. The bounce and the false-vacuum configurations must be compared using
the same affine boundary data. This means that the same affine radial
representative is used as a function of the physical radius:
\begin{equation}\label{eq:same_affine_representative}
	s_b(r)=S(\rho_b(r)),
	\qquad
	s_{\rm fv}(r)=S(\rho_{\rm fv}(r)),
\end{equation}
with the same function \(S\). Using Eq.\ \eqref{eq:c_from_affine_map}, one
then obtains, at the matched physical boundary,
\begin{equation}\label{eq:c_boundary_same_S}
	c_b(\bar r_b)
	=
	-
	\frac{S(\bar\rho)}
	{S_{,\rho}(\bar\rho)\dot\rho_b(\bar r_b)},
	\qquad
	c_{\rm fv}(\bar r_{\rm fv})
	=
	-
	\frac{S(\bar\rho)}
	{S_{,\rho}(\bar\rho)\dot\rho_{\rm fv}(\bar r_{\rm fv})}.
\end{equation}
For the Minkowski false-vacuum geometry one has
\[
\rho_{\rm fv}(r)=r,
\qquad
\dot\rho_{\rm fv}(r)=1,
\]
and therefore
\begin{equation}\label{eq:cb_minus_cfv_general}
	c_b(\bar r_b)-c_{\rm fv}(\bar r_{\rm fv})
	=
	\frac{S(\bar\rho)}{S_{,\rho}(\bar\rho)}
	\left(
	1-\frac{1}{\dot\rho_b(\bar r_b)}
	\right).
\end{equation}
For an asymptotically Minkowskian affine representative,
\begin{equation}\label{eq:asymptotic_affine_representative}
	\frac{S(\bar\rho)}{S_{,\rho}(\bar\rho)}
	\sim
	\bar\rho
	\qquad
	(\bar\rho\to\infty).
\end{equation}
Hence
\begin{equation}\label{eq:cb_minus_cfv_falloff_general}
	c_b(\bar r_b)-c_{\rm fv}(\bar r_{\rm fv})
	=
	O\left[
	\bar\rho\,
	\bigl(\dot\rho_b(\bar r_b)-1\bigr)
	\right].
\end{equation}

Using the asymptotic behavior derived in Sec.~\ref{asymptotic}, this
difference vanishes at infinity. In the massive false-vacuum case,
Eq.\ \eqref{eq:dotrhomass} gives
\begin{equation}
	c_b-c_{\rm fv}
	=
	O(\bar\rho^3e^{-2\mu\bar\rho})
	\longrightarrow0 ,
\end{equation}
whereas in the massless case Eq.\ \eqref{eq:dotrhomassless} gives
\begin{equation}
	c_b-c_{\rm fv}
	=
	O(\bar\rho^{-3})
	\longrightarrow0 .
\end{equation}
Therefore, in both cases,
\begin{equation}\label{eq:c_matching_infinity}
	c_b(\bar r_b)-c_{\rm fv}(\bar r_{\rm fv})
	\longrightarrow0
	\qquad
	(\bar\rho\to\infty).
\end{equation}

Thus, the affine matching specifies the asymptotic boundary data of the STEGR problem: the bounce and the false
vacuum have to be compared within the same affine representative. Comparing
them in different asymptotic affine sectors would correspond to comparing different STEGR boundary-value problems.

Having determined the non-vanishing components of $Q_\mu$ and $\tilde Q_\mu$, Eq.\,\eqref{QQtraces}, together with the behavior of the function $c(r)$ at the endpoints (see Eqs.\,\eqref{eq:c_origin_expansion} and \eqref{eq:c_matching_infinity}) in both the massive and the massless cases,  we can now proceed with the calculation of the contribution to the tunneling exponent $B$ coming from the boundary term in \eqref{eq:VV}. This is the objective of the next section.

\subsection{Tunneling exponent}
\label{sec:tunneling}

We now have all the ingredients to evaluate the STEGR action
\eqref{STEGRbounceaction} at the bounce solution. Using the geometric identity
\eqref{eq:GR_STEGR}, the action can be written as the GR one
plus the total-derivative contribution of the two non-metricity traces.
For the bounce one obtains
\begin{align}\label{ESTEGR2}
	S_b^{^{\rm STEGR}}
	&=
	\int_{\mathbb R^4} d^4x\sqrt{g}
	\left[
	-\frac{Q}{2\kappa}
	+\frac12 g^{\mu\nu}\nabla_\mu\phi\nabla_\nu\phi
	+V(\phi)
	\right]_{\phi=\phi_b,\,\rho=\rho_b}
	\nonumber
	\\
	&=
	S_b^{^{\rm GR}}
	-\frac{1}{2\kappa}
	\int_{\mathbb R^4} d^4x\sqrt{g}\,
	\gr\nabla_\mu\left(Q^\mu-\tilde Q^\mu\right)
	\Big|_{\phi=\phi_b,\,\rho=\rho_b},
\end{align}
where \((\phi_b,\rho_b)\) is the bounce solution, the same as in GR, and
\(S_b^{^{\rm GR}}\) is the GR bounce action. A similar relation holds for the
false vacuum configuration.

To calculate the boundary term in the second line of \eqref{ESTEGR2}, we
first evaluate it on a finite radial domain. More precisely, we denote by \(\mathcal M_{\bar r}\) the four-sphere of $\mathbb R^4$ of radius $\bar r$. Its boundary $\partial \mathcal M_{\bar r}$ is the
hypersurface $r=\bar r$, i.e.\,\,the three-sphere $S^3$ of radius $\bar\rho=\rho(\bar r)$. As stressed in the previous Section (see \eqref{eq:boundary}), for the bounce and the false vacuum we choose two different coordinate positions, $\bar r_b$ and  $\bar r_{\rm fv}$ respectively, such that
$
\rho_b(\bar r_b)=\rho_{\rm fv}(\bar r_{\rm fv})=\bar\rho$. Since, for the configurations considered here, \(\bar\rho\to\infty\) when the
outer radial boundary $\bar r$ is sent to infinity, the asymptotic limit is taken by
sending this common physical radius \(\bar\rho\) to infinity.

We therefore define
\begin{equation}\label{eq:IQ}
	\mathcal I_{\bar r}[\rho,c]
	\equiv
	\int_{\mathcal M_{\bar r}} d^4x\sqrt{g}\,
	\gr\nabla_\mu\left(Q^\mu-\tilde Q^\mu\right).
\end{equation}
As already seen, only the radial component of
\(Q_\mu\) and \(\tilde Q_\mu\) is non-vanishing. Inserting \eqref{QQtraces}
in \eqref{eq:IQ}, we get
\begin{align}\label{boundarygeneral}
	\mathcal I_{\bar r}[\rho,c]
	=
	\int_{\mathcal M_{\bar r}} d^4x\,
	\partial_\mu
	\left[
	\sqrt{g}\,g^{\mu\nu}
	\left(Q_\nu-\tilde Q_\nu\right)
	\right]
	=
	6\pi^2
	\left[
	\rho
	\left(
	2\rho\dot\rho+\frac{\rho^2}{c}+c
	\right)
	\right]_{0}^{\bar r}.
\end{align}
It is useful to introduce the shorthand
\begin{equation}\label{eq:Fboundary}
	\mathcal F_Q[\rho,c]
	\equiv
	\rho
	\left(
	2\rho\dot\rho+\frac{\rho^2}{c}+c
	\right).
\end{equation}
Then the STEGR surface contribution is controlled by the endpoint values of
\(\mathcal F_Q\).

We first analyze the false vacuum configuration. As we know from Section~\ref{asymptotic} (see \eqref{eq:falsevacuumdef} and comments below), in this case $\phi_{\rm fv}=0$, $\rho_{\rm fv}(r)=r$. Taking the the corresponding
regular affine map
introduced in \eqref{eq:affine_coordinates_s}, we have
\(s_{\rm fv}(r)=r\). Inserting this relation in \eqref{eq:c_from_affine_map} have $c_{\rm fv}(r)=-r$, and then
\(c_{\rm fv}(r)=-\rho_{\rm fv}(r)\). Therefore
\begin{equation}
	\mathcal F_Q[\rho_{\rm fv},c_{\rm fv}]
	=
	\rho_{\rm fv}
	\left(
	2\rho_{\rm fv}
	-\rho_{\rm fv}
	-\rho_{\rm fv}
	\right)
	=
	0 .
\end{equation}
Thus the STEGR surface contribution vanishes identically on the false vacuum
configuration, for any value of the outer radius:
\begin{equation}\label{eq:Ifv_zero}
	\mathcal I_{\bar r}[\rho_{\rm fv},c_{\rm fv}]=0 .
\end{equation}

Let us now consider the boundary term \eqref{boundarygeneral} for the bounce. At the regular endpoint
\(r\to0\), the metric field behave as in Eq.\ \eqref{rto0}, i.e.\,\,$	\rho_b(r)=r+O(r^3)$.
Moreover, regularity of the affine map gives the endpoint behavior $c_b(r)=-r+O(r^2)$,
as derived in Eq.\ \eqref{eq:c_origin_expansion} from the pure-gauge
representation \eqref{eq:c_from_affine_map}. Substituting these expressions
into Eq.\ \eqref{eq:Fboundary}, one finds $	\mathcal F_Q[\rho_b,c_b]
=
O(r^4)$
and therefore
\begin{equation}\label{eq:bounce_origin_zero}
	\lim_{r\to0}\mathcal F_Q[\rho_b,c_b]=0 .
\end{equation}

We next analyze the asymptotic endpoint. As discussed in
Sec.~\ref{affineconnectioncalc}, the bounce and the false vacuum have to be
compared at the same physical boundary and within the same asymptotic affine
representative. Therefore, in the following we use the matched boundary
prescription introduced there. In particular, at the outer endpoint one has
\begin{equation}
	c_{\rm fv}(\bar r_{\rm fv})=-\bar\rho,
	\qquad
	c_b(\bar r_b)=c_{\rm fv}(\bar r_{\rm fv})+\delta c(\bar\rho)
	=-\bar\rho+\delta c(\bar\rho),
	\qquad
	\delta c(\bar\rho)\to0 ,
\end{equation}
as follows from Eqs.\ \eqref{eq:same_affine_representative} and
\eqref{eq:c_matching_infinity}. Moreover, from the asymptotic analysis of
Sec.~\ref{asymptotic}, in particular from
Eqs.\ \eqref{eq:weighted_rho_falloff_massive} and
\eqref{eq:weighted_rho_falloff_massless}, one has in both the massive and
massless cases $\bar\rho^2\left(\dot\rho_b(\bar r_b)-1\right)\longrightarrow0$ for $\bar \rho \to \infty$.
Defining
$
\epsilon_b(\bar\rho)\equiv \dot\rho_b(\bar r_b)-1$, we can write \eqref{eq:Fboundary} at the asymptotic boundary as
\begin{align}
	\mathcal F_Q[\rho_b,c_b]\Big|_{\bar r_b}
	=
	\bar\rho
	\left[
	2\bar\rho(1+\epsilon_b)
	+
	\frac{\bar\rho^2}{-\bar\rho+\delta c}
	-\bar\rho+\delta c
	\right]
	=
	2\bar\rho^2\epsilon_b
	+
	O(\delta c^2).
\end{align}
Since \(\bar\rho^2\epsilon_b\to0\) and \(\delta c\to0\), it follows that $	\lim_{\bar\rho\to\infty}
\mathcal F_Q[\rho_b,c_b]\Big|_{\bar r_b}
=
0 $.
Therefore the STEGR surface contribution also vanishes on the bounce:
\begin{equation}\label{eq:Ib_zero}
	\lim_{\bar\rho\to\infty}
	\mathcal I_{\bar r_b}[\rho_b,c_b]=0 .
\end{equation}

	Equations \eqref{eq:Ifv_zero} and \eqref{eq:Ib_zero} are the main result of
	the present section. As already stressed, at the leading order of the semiclassical expansion the decay rate $\Gamma^{^{\rm STEGR}}$ is given by
	\begin{equation}
	\Gamma^{^{\rm STEGR}}\sim e^{-B^{^{\rm STEGR}}} =e^{-(S_b^{^{\rm STEGR}}-S_{\rm fv}^{^{\rm STEGR}})}\,,
	\end{equation} 
	where (see \eqref{ESTEGR2})
	\begin{equation}
		S^{^{\rm STEGR}}=S^{^{\rm GR}}+{\rm \vv boundary\,\,term"}\,.
	\end{equation} 
	The analysis performed above shows that the
	contribution to $B^{^{\rm STEGR}}$ from the \vv boundary term" actually vanishes. This result follows from the explicit endpoint calculation, and is one of the main findings of this work: at the leading order of the semiclassical expansion, the false vacuum decay rate $\Gamma^{^{\rm STEGR}}$ for the Minkowski-to-AdS decay coincides with $\Gamma^{^{\rm GR}}$. 
	
	In the next Section we will perform a similar analysis for the teleparallel equivalent to GR (TEGR), calculating the contribution to the tunneling exponent $B$ coming from the boundary term.

\section{Vacuum stability in TEGR}
\label{vacdecayTEGR}
Here, we perform the vacuum stability analysis in the TEGR framework, considering the TEGR action \eqref{eq:TEGR_lagrangian} in the
Euclidean setting. The (Euclidean) torsion scalar, that we denote with $T_E$, is constructed using the
tetrad and the corresponding flat spin connection as in Eq.\ \eqref{eq: torsion}. The action for the scalar field minimally coupled to gravity, together with the $\frac{T_E}{2\kappa}$ term, is (\(e_E=\sqrt{g_E}\))
\begin{equation}\label{TEGRbounceaction}
S_E[\phi,e,\omega]
=
\int_{\mathbb R^4}d^4x\,e_E
\left[
\frac{T_E}{2\kappa}
+\frac12 g_E^{\mu\nu}\nabla_\mu\phi\nabla_\nu\phi
+V(\phi)
\right].
\end{equation}
In the following we work in Euclidean signature and drop the subscript \(E\).

The geometric identity \eqref{eq:GR_TEGR}
holds also in the Euclidean setting, since it is a purely geometric
relation involving the metric-compatible affine connection
$\Gamma^\alpha{}_{\mu\nu}
=
\gr{\Gamma}^\alpha{}_{\mu\nu}
+
K^\alpha{}_{\mu\nu}$.
Moreover, since \(\phi\) is a scalar field, its covariant derivative reduces
to the ordinary partial derivative,
$\nabla_\mu\phi=\partial_\mu\phi$.
Therefore, as in the STEGR case, the matter sector does not distinguish
between the Levi-Civita and the teleparallel connection.

	As a consequence the TEGR action
	\eqref{TEGRbounceaction} is dynamically equivalent to the GR action
	\eqref{gaction} \cite{BeltranJimenez:2020sih,Oshita:2017nhn}. The TEGR and GR actions differ
	by a total derivative, leading to the same field equations.
	Therefore, the \(O(4)\)-symmetric bounce solution is the same as in GR (obeying to the same equations \eqref{gequation} and \eqref{eq:rhodot} and
	boundary conditions \eqref{bcond})
	\begin{equation}
		\{\phi_b(r),\rho_b(r)\}_{\rm TEGR}
		=
		\{\phi_b(r),\rho_b(r)\}_{\rm GR}.
	\end{equation}

	The bulk equivalence does not by itself guarantee equality of the on-shell actions: we do not know a priori whether $S_b^{^{\rm TEGR}}=S_b^{^{\rm GR}}$ and $S_{\rm fv}^{^{\rm TEGR}}=S_{\rm fv}^{^{\rm GR}}$, that in turn would imply $B^{^{\rm TEGR}}=B^{^{\rm GR}}$. In fact, their on-shell values may differ
	by endpoint contributions. This is particularly relevant for the non-compact
	bounce describing the decay of a Minkowski false vacuum. As for the STEGR  case,  we have to perform an endpoint analysis for the bounce and false vacuum solutions, checking whether the boundary term gives (or not) a non-trivial contribution to $B^{^{\rm TEGR}}$. Only after that, it will be possible to determine the contribution $B^{^{\rm TEGR}}_{\rm boundary}$ to $B^{^{\rm TEGR}}$.

To perform this analysis, we have to derive the most general \(O(4)\)-invariant flat and metric-compatible affine connection, since the latter is needed to evaluate the TEGR boundary
term. We do it in the next section. 

\subsection{\texorpdfstring{\(\boldsymbol{O(4)}\)}{O(4)}-invariant tetrad and spin connection}
\label{O(4)tetradcalculation}

We determine now the \(O(4)\)-invariant
flat and metric-compatible affine connection. In the TEGR formulation of gravity this is most
naturally done in the tetrad language. The affine connection is obtained from
the tetrad postulate \eqref{eq: tetrad postulate}, once a tetrad
\(e^a{}_\mu\) and its corresponding flat spin connection
\(\omega^a{}_{b\mu}\) have been specified. We have then to formulate the theory in terms of a tetrad and of the (corresponding) spin connection with $O(4)$ invariance. The \(O(4)\)-invariant line element for the bounce is given in \eqref{metric}, and a simple tetrad
reproducing it is the diagonal one,
\begin{equation}\label{diagonale}
	e^a{}_{\mu}
	=
	{\rm diag}
	\left(
	1,\rho(r),\rho(r)\sin\chi,\rho(r)\sin\chi\sin\theta
	\right).
\end{equation}
In order to determine the corresponding spin connection \(\omega^a{}_{b\mu}\) we now follow the usual pattern. First we determine the $SO(4)$ matrix $\Lambda^a{}_b(x)$ that allows to obtain the proper tetrad from the diagonal one \cite{Hohmann:2019nat,Hohmann:2019fvf}, then we calculate \(\omega^a{}_{b\mu}\) from Eq.\ \eqref{eq: omega inerziale}. As it is well-known, going from \eqref{diagonale} to the corresponding flat tetrad $(e^a{}_{\mu})_{\rm flat}
=
{\rm diag}
\left(
1,r,r\sin\chi,r\sin\chi\sin\theta
\right)$, the matrix $\Lambda^a{}_b(x)$ is the matrix that connects $(e^a{}_{\mu})_{\rm flat}$ in spherical coordinates with the corresponding $(e^a{}_{\mu})_{\rm flat}$ in Cartesian coordinate. Namely,
\begin{align}\label{Lambda}
	\Lambda^a{}_b(\chi,\theta,\varphi)=
	\begin{pmatrix}
		\cos \chi & -\sin \chi & 0 & 0 \\
		\sin \chi \cos \theta & \cos \chi \cos \theta & -\sin \theta & 0 \\
		\sin \chi \sin \theta \cos \varphi & \cos \chi \sin \theta \cos \varphi & \cos \theta \cos \varphi & -\sin \varphi \\
		\sin \chi \sin \theta \sin \varphi & \cos \chi \sin \theta \sin \varphi & \cos \theta \sin \varphi & \cos \varphi
	\end{pmatrix}.
\end{align}
From \eqref{eq: omega inerziale}, it is now immediate to calculate the spin connection \(\omega^a{}_{b\mu}\) corresponding to the tetrad \eqref{diagonale}. The non-vanishing
components are (\(\omega^a{}_{br}=0\))
\begin{align}\label{spin connection}
\omega^a{}_{b\chi}
&=
\begin{pmatrix}
0 & -1 & 0 & 0 \\
1 & 0 & 0 & 0 \\
0 & 0 & 0 & 0 \\
0 & 0 & 0 & 0
\end{pmatrix},
\qquad
\omega^a{}_{b\theta}
=
\begin{pmatrix}
0 & 0 & -\sin \chi & 0 \\
0 & 0 & -\cos \chi & 0 \\
\sin \chi & \cos \chi & 0 & 0 \\
0 & 0 & 0 & 0
\end{pmatrix},
\nonumber
\\
\omega^a{}_{b\varphi}
&=
\begin{pmatrix}
0 & 0 & 0 & -\sin \chi \sin \theta \\
0 & 0 & 0 & -\cos \chi \sin \theta \\
0 & 0 & 0 & -\cos \theta \\
\sin \chi \sin \theta & \cos \chi \sin \theta & \cos \theta & 0
\end{pmatrix}.
\end{align}
Finally, the \(O(4)\)-invariant flat and
metric-compatible affine connection is obtained from the tetrad postulate
\eqref{eq: tetrad postulate}. We have

\begin{align}\label{connectionTflat}
\Gamma^r{}_{\mu\nu}
&=
\begin{pmatrix}
0 & 0 & 0 & 0 \\
0 & -\rho(r) & 0 & 0 \\
0 & 0 & -\rho(r)\sin^2\chi & 0 \\
0 & 0 & 0 & -\rho(r)\sin^2\chi \sin^2\theta
\end{pmatrix},
\nonumber
\\
\Gamma^\chi{}_{\mu\nu}
&=
\begin{pmatrix}
0 &\frac{1}{\rho(r)} & 0 & 0 \\
\frac{\dot \rho(r)}{\rho(r)} & 0 & 0 & 0 \\
0 & 0 & -\sin\chi \cos \chi & 0 \\
0 & 0 & 0 & -\sin\chi \cos \chi\sin^2\theta
\end{pmatrix},
\nonumber
\\
\Gamma^\theta{}_{\mu\nu}
&=
\begin{pmatrix}
0 & 0 & \frac{1}{\rho(r)} & 0 \\
0 & 0 & \cot \chi & 0 \\
\frac{\dot \rho(r)}{\rho(r)} & \cot \chi & 0 & 0 \\
0 & 0 & 0 & -\sin\theta \cos\theta
\end{pmatrix},
\qquad
\Gamma^\varphi{}_{\mu\nu}
=
\begin{pmatrix}
0 & 0 & 0 & \frac{1}{\rho(r)} \\
0 & 0 & 0 & \cot \chi \\
0 & 0 & 0 & \cot \theta \\
\frac{\dot \rho(r)}{\rho(r)} & \cot \chi & \cot \theta & 0
\end{pmatrix}.
\end{align}
In passing, we note that the affine connection \eqref{connectionTflat} could also been obtained imposing directly \(O(4)\)
invariance to the components $\Gamma^\alpha{}_{\mu\nu}$ (as done in Section \ref{affineconnectioncalc} for the STEGR case, see Eq.\,\ref{O4}) together with the teleparallel
metric conditions $R^\rho{}_{\sigma\mu\nu}=0$ and $Q_{\alpha\mu\nu}=0$. 

Using Eq.\ \eqref{connectionTflat}, the torsion scalar \eqref{eq: T tegr}
becomes
\begin{align}\label{O(4)Tscalar}
T
=
-\frac{6}{\rho(r)^2}
\left(
\dot \rho(r)-1
\right)^2 .
\end{align}
Moreover, the torsion vector \(T_\mu\), defined in Eq.\ \eqref{eq:Ta}, has only
one non-vanishing component,
\begin{align}\label{O(4)Tvect}
T_r
=
-\frac{3}{\rho(r)}
\left(
\dot \rho(r)-1
\right).
\end{align}

Having determined in Eq.\,\eqref{O(4)Tvect} the non-vanishing component of $T_\mu$, in the next subsection we
will use the asymptotic analysis of
Section \ref{asymptotic} to evaluate the endpoint behavior of boundary term relating TEGR to GR, and then compute
the corresponding contribution to the tunneling exponent \(B\).

\subsection{Tunneling exponent}
\label{Sec:TEGR Tunnel}

We now evaluate the TEGR action \eqref{TEGRbounceaction} at the bounce
solution. Using the geometric identity \eqref{eq:GR_TEGR}, the action can be
written as the GR one plus the total-derivative contribution of the torsion
vector. For the bounce one obtains
\begin{align}\label{ETEGR2}
	S_b^{^{\rm TEGR}}
	&=
	\int_{\mathbb R^4} d^4x\,e
	\left[
	\frac{T}{2\kappa}
	+\frac12 g^{\mu\nu}\nabla_\mu\phi\nabla_\mu\phi
	+V(\phi)
	\right]_{\phi=\phi_b,\,\rho=\rho_b}
	\nonumber
	\\
	&=
	S_b^{^{\rm GR}}
	-\frac{1}{2\kappa}
	\int_{\mathbb R^4} d^4x\,e\,
	2\gr\nabla_\mu T^\mu
	\Big|_{\phi=\phi_b,\,\rho=\rho_b},
\end{align}
where \((\phi_b,\rho_b)\) is the bounce solution, the same as in GR, and
\(S_b^{^{\rm GR}}\) is the GR bounce action. A similar relation holds for the
false vacuum configuration.

Following the analysis of Subsection \ref{sec:tunneling}, to evaluate the boundary term of \eqref{ETEGR2}, we first
restrict the integration to the finite four-dimensional domain
\(\mathcal M_{\bar r}\subset\mathbb R^4\), defined by \(0\le r\le\bar r\).
Its boundary \(\partial\mathcal M_{\bar r}\) is the hypersurface
\(r=\bar r\), namely a three-sphere with physical radius
\(\rho(\bar r)\). Following the boundary prescription discussed in
Sec.~\ref{affineconnectioncalc}, the bounce and the false vacuum configurations
are then compared at the same physical boundary, choosing two coordinate positions \(\bar r_b\) and \(\bar r_{\rm fv}\), respectively, so that
$
	\rho_b(\bar r_b)=\rho_{\rm fv}(\bar r_{\rm fv})=\bar\rho$.
The asymptotic limit is finally taken by sending the physical radius
\(\bar\rho\) to infinity.

We therefore define
\begin{equation}\label{eq:JT}
	\mathcal J_{\bar r}[\rho]
	\equiv
	\int_{\mathcal M_{\bar r}} d^4x\,e\,
	\gr\nabla_\mu T^\mu .
\end{equation}
We know that only the radial component of $T_\mu$ is non-vanishing. Inserting Eq.\ \eqref{O(4)Tvect} in \eqref{eq:JT} we obtain
\begin{align}\label{boundarygeneralTEGR}
	\mathcal J_{\bar r}[\rho]
	=
	\int_{\mathcal M_{\bar r}} d^4x\,
	\partial_\mu
	\left(
	e\,g^{\mu\nu}T_\nu
	\right)
	=
	-6\pi^2
	\left[
	\rho^2(\dot\rho-1)
	\right]_{0}^{\bar r}.
\end{align}
It is useful to introduce the shorthand
\begin{equation}\label{eq:FTboundary}
	\mathcal F_T[\rho]
	\equiv
	\rho^2(\dot\rho-1).
\end{equation}

Let us first consider the false vacuum configuration. For the
Minkowski false vacuum, $\phi_{\rm fv}=0$ and $\rho_{\rm fv}(r)=r$, we get
\begin{equation}
	\mathcal F_T[\rho_{\rm fv}]
	=
	\rho_{\rm fv}^2
	\left(
	\dot\rho_{\rm fv}-1
	\right)
	=
	0 ,
\end{equation}
so that the TEGR surface contribution vanishes identically on the false vacuum
configuration, for any value of the outer radius:
\begin{equation}\label{eq:JfvTEGR_zero}
	\mathcal J_{\bar r}[\rho_{\rm fv}]=0 .
\end{equation}

Let us now consider the bounce. At the regular endpoint \(r\to0\), the metric
 behaves as in Eq.\ \eqref{rto0}. i.e.\,\,$\rho_b(r)=r+O(r^3)$.
Substituting this expression into Eq.\ \eqref{eq:FTboundary}, one obtains
\begin{equation}
	\mathcal F_T[\rho_b]
	=
	O(r^4)
	\qquad
	(r\to0),
\end{equation}
and therefore
\begin{equation}\label{eq:bounce_TEGR_origin_zero}
	\lim_{r\to0}\mathcal F_T[\rho_b]=0 .
\end{equation}
It remains to analyze the asymptotic endpoint. Evaluating the function $\mathcal F_T$ in \eqref{eq:FTboundary} at $r=\bar r_b $, we get
\begin{equation}
	\mathcal F_T[\rho_b]\Big|_{\bar r_b}
	=
	\bar\rho^2
	\left[
	\dot\rho_b(\bar r_b)-1
	\right].
\end{equation}
The leading asymptotic condition \(\dot\rho_b\to1\) is not sufficient by
itself to guarantee the vanishing of this endpoint term, because it is
multiplied by \(\bar\rho^2\). What is needed is the stronger weighted falloff
derived in Sec.~\ref{asymptotic}. In particular, from
Eqs.\ \eqref{eq:weighted_rho_falloff_massive} and
\eqref{eq:weighted_rho_falloff_massless}, one has in both the massive and
massless cases $\bar\rho^2
\left[
\dot\rho_b(\bar r_b)-1
\right]
\longrightarrow0$.
It follows that $	\lim_{\bar\rho\to\infty}
\mathcal F_T[\rho_b]\Big|_{\bar r_b}
=
0$, and then the TEGR surface contribution also vanishes on the bounce:
\begin{equation}\label{eq:JbTEGR_zero}
	\lim_{\bar\rho\to\infty}
	\mathcal J_{\bar r_b}[\rho_b]=0 .
\end{equation}

Eqs.\,\eqref{eq:JfvTEGR_zero} and \eqref{eq:JbTEGR_zero} are the main
result of the present section. The endpoint analysis shows that the
contribution coming from the TEGR boundary term vanishes separately on the
false vacuum and on the bounce, for the same boundary prescription used in the
STEGR analysis. Thus, the total derivative relating TEGR and GR gives no
contribution to the tunneling exponent $B^{^{\rm TEGR}}$.

This conclusion is not simply a restatement of the bulk equivalence with GR.
The dynamical equivalence of GR and TEGR guarantees that both theories share the same \(O(4)\)-symmetric saddle, but it does not by itself determine the endpoint value of $\hat{B}$ . In the TEGR case, the endpoint value of the
torsional boundary current is fixed by
the explicit computation of the torsion vector using the covariant
\(O(4)\)-invariant tetrad-spin-connection pair. This gives
Eq.\ \eqref{O(4)Tvect} and hence the endpoint function
\eqref{eq:FTboundary}. An indiscriminate use of the diagonal spherical tetrad
with vanishing spin connection would instead retain spurious inertial torsion
and could contaminate the surface calculation.

Therefore, our result states that, at the leading order of the semiclassical expansion, the decay rate $\Gamma^{^{\rm TEGR}}$ is given by
\begin{equation}
	\Gamma^{^{\rm TEGR}}\sim e^{-B^{^{\rm TEGR}}} =e^{-(S_b^{^{\rm TEGR}}-S_{\rm fv}^{^{\rm TEGR}})}\,,
\end{equation} 
where 
\begin{equation}
	S^{^{\rm TEGR}}=S^{^{\rm GR}}+{\rm \vv boundary\,\,term"}\,.
\end{equation}

\section{Conclusions and outlook}
In this work, we investigated the false vacuum decay within the framework of Geometric Trinity of Gravity, comparing GR with its teleparallel and symmetric teleparallel equivalents, TEGR and STEGR. At first sight, the equality of the corresponding decay rates might appear to follow immediately from their classical equivalence. In fact, the three formulations lead to the same bulk field equations and therefore admit the same $O(4)$-symmetric Coleman-De Luccia bounce. Nevertheless, this is not sufficient to establish that the tunneling rate $\Gamma \sim e^{-B}$ at the leading order ($B$ tunneling exponent) is the same in the three formulations. Actually the boundary terms and the
non-trivial inner and asymptotic endpoints behaviors of the bounce may influence the leading semiclassical contribution $e^{-B}$ of the decay rate.
The essential point is that the tunneling exponent $B$ is determined by the subtraction between the Euclidean bounce action and the false vacuum one,
\[
B=S_E[\text{bounce}]-S_E[\text{false vacuum}],
\]
rather than by the bulk equations of motion alone. Although the GR, TEGR, and STEGR actions differ only by a boundary term, the latter could a prori produce finite endpoint contributions when evaluated on a non-compact Euclidean configuration, like the bounce. Their contribution depends on the regularity of the fields at the origin, the asymptotic falloff of the bounce, the prescription used to match the bounce to the false vacuum background, and, in both teleparallel and symmetric teleparallel formulations, the choice of affine or inertial data. Consequently, even if the three formulations of gravity share the same bounce solution, this does not guarantee the same semiclassical weight to the decay rate.

In order to analyze the contribution of the boundary terms of TEGR and STEGR, we compared the Euclidean bounce action and the false vacuum one, using the same physical regulator for the perform the asymptotic analysis at $r \to \infty$. The bounce and false vacuum geometries were matched on a three-sphere of fixed physical radius, and the background subtraction was performed before removing the cutoff. Then, we evaluated explicitly the torsion and non-metricity boundary currents at both the regular origin and the asymptotic endpoint.

In STEGR, this analysis requires the construction of the most general flat, torsionless, and \(O(4)\)-invariant affine connection. Nevertheless, even if we impose these conditions, the affine connection still contains a residual radial function $c(r)$.  This function does not introduce an additional bulk degree of freedom or modify the bounce solution, but it enters the non-metricity boundary current, affecting the on-shell action. In fact, the non-metricity boundary term is controlled by the function
\[
\mathcal F_Q[\rho,c]
\equiv
\rho
\left(
2\rho\dot\rho+\frac{\rho^2}{c}+c
\right).
\]
The regularity of the affine map fixes the behavior of $c(r)$ near the origin, while a meaningful comparison with the false vacuum requires the bounce and false vacuum configurations to belong to the same asymptotic affine sector, from which $c_b-c_{\rm fv} \to 0$ for $r \to \infty$. Under these conditions, using the regularity of the bounce at the origin and the asymptotic falloff $
\rho^2(\dot{\rho}-1)\longrightarrow0
$ of the bounce derived in Sec.~\ref{sec:asymptotic-bounce}, we were able to show that the STEGR boundary contribution vanishes at both regular origin and asymptotic endpoint.

The TEGR calculation involves a distinct subtlety. The torsional surface term must be evaluated using a covariant \(O(4)\)-invariant tetrad and spin connection pair. A diagonal tetrad in spherical coordinates, combined with a vanishing spin connection, would retain spurious inertial torsion and could contaminate the value of the action, even though it reproduces the correct bulk equations. With the appropriate inertial spin connection, or equivalently with the corresponding non-diagonal tetrad in the Weitzenböck gauge, the TEGR boundary contribution is controlled by
$\mathcal F_T[\rho]=\rho^2(\dot{\rho}-1)$.
The regularity of the bounce at the origin makes this quantity vanish at $r=0$. At infinity, the leading condition $\dot{\rho}\to1$ is not sufficient because it is multiplied by the growing factor $\rho^2$. The stronger result
$
\rho^2(\dot{\rho}-1)\longrightarrow0
$ ensures the vanishing of the TEGR contribution at the asymptotic endpoint.

The endpoint analysis establishes that, for regular teleparallel representatives and consistently matched metric and affine boundary data, the relevant surface terms vanish separately on the bounce and on the false vacuum configuration. Thus, in this work, we proved that 
$B_{\mathrm{GR}}=B_{\mathrm{STEGR}}=B_{\mathrm{TEGR}}$
for the $O(4)$-symmetric decay of a Minkowski false vacuum into an AdS true vacuum. 

Equality $B_{\mathrm{GR}}=B_{\mathrm{STEGR}}=B_{\mathrm{TEGR}}$ is the outcome of an explicit action-level calculation that shows that the distinct torsion and non-metricity boundary currents do not leave a residual contribution to the tunneling exponent $B$. This does not depend on the dynamical equivalence among GR, TEGR, and STEGR.

The role of these currents should also be distinguished from that of the Gibbons-Hawking-York term. The latter is added to the EH action to obtain a well-posed Dirichlet variational problem. On the other hand, the TEGR and STEGR total derivatives arise from the geometric identities relating the curvature, torsion, and non-metricity scalars. The cancellation of the GHY contribution in the GR tunneling exponent might induce the expectation of a similar result for the teleparallel and symmetric teleparallel boundary terms, though as shown in the present work this cancellation is far from being trivial. In particular, the STEGR current depends on independent affine connection, whereas the GHY term is determined by the induced metric and its extrinsic curvature.

Our result provides a non-trivial semiclassical consistency test of the Geometric Trinity. It shows that the classical equivalence extends to the leading tunnelling exponent for the specific matter coupling, boundary conditions, asymptotic sector, and class of regular configurations studied here. It should not, however, be interpreted as a general proof of quantum equivalence. In particular, the fluctuation prefactor, the functional measure, and configurations with different asymptotics, internal boundaries, or affine boundary data may require a separate analysis.

Several extensions naturally follow from these observations. It would be particularly interesting to study compact bounces and other false vacuum geometries, for which the endpoint structure differs from the Minkowski false vacuum case, and the boundary currents need not vanish in the same way. More general metric-affine theories, including matter sourcing hypermomentum through a direct coupling to the independent connection, may also distinguish between the three geometric descriptions. Another direction is provided by extensions such as \(f(R)\), \(f(Q-B)\), and \(f(-T-\hat B)\) , in which the relevant boundary combinations enter non-linearly and the classical equivalence with GR is generically lost. Finally, extending the comparison to the one-loop fluctuation determinants, or developing a Hamiltonian formulation of the Coleman-De Luccia process, would clarify how far the equivalence of the Geometric Trinity persists beyond the leading semiclassical exponent.

\begin{acknowledgments}
We  thank our colleagues  Pietro Conzinu, Riccardo Gandolfo, Damianos Iosifidis, and Arcangelo Pernace   for the useful discussions, criticisms and comments. 
We acknowledge the support of Istituto Nazionale di Fisica Nucleare (INFN), Sezione di Napoli, \textit{Iniziativa Specifica} QGSKY. This publication is based upon work from COST Action CA21136 - ``Addressing observational tensions in cosmology with systematics and fundamental physics (CosmoVerse)'', supported by COST (European Cooperation in Science and Technology).

\end{acknowledgments}

\appendix
\section{The \texorpdfstring{$O(4)$}{O(4)} symmetry}
\label{Symmetry}
The generators of the $SO(4)$ group are given by the differential operators
\begin{align}  \label{generators}
J_{\mu\nu}=x_\mu \partial_\nu-x_\nu\partial_\mu\,.
\end{align}
In spherical coordinates, the generators \eqref{generators} are explicitly
\begin{align}
J_{12}&=\partial_\varphi\,, \qquad J_{13}=-\cos \varphi \,\partial_\theta+\frac{\sin\varphi}{\tan \theta}\,\partial\varphi\,,  \nn \\
J_{14}&=-\sin \theta \cos \varphi\, \partial_\chi-\cot \chi \cos \theta \cos \varphi\,\partial_\theta+\frac{\sin \varphi}{\tan \chi \sin \theta}\,\partial_\phi\,, \quad \,\, J_{23}=-\sin \varphi \,\partial_\theta-\frac{\cos\varphi}{\tan \theta}\,\partial_\varphi\,,\nn \\
J_{24}&=-\sin \theta \sin \varphi\, \partial_\chi-\cot \chi \cos \theta \sin \varphi\,\partial_\theta-\frac{\cos \varphi}{\tan \chi \sin \theta}\,\partial_\varphi\,, \nn \\
J_{34}&=-\cos \theta\,\partial_\chi+ \cot \chi\sin \theta\,\partial_\theta\,,
\end{align}
so that the corresponding Killing vectors are respectively
\begin{align}\label{Killing}
X^\mu_1&=\left( 0,0,0,1 \right)\,, \qquad X^\mu_2=\left( 0,0,-\cos \phi,\frac{\sin\varphi}{\tan\theta} \right)\,,  \nn \\
X^\mu_3&=\left( 0,-\sin \theta \cos \varphi,-\cot \chi \cos \theta \cos \varphi,\frac{\sin\varphi}{\tan\chi \sin\theta} \right)\,,\quad X^\mu_4=\left( 0,0,-\sin \varphi,-\frac{\cos\varphi}{\tan\theta} \right)\,, \nn \\
X^\mu_5&=\left( 0,-\sin \theta \sin \varphi,-\cot \chi \cos \theta \sin \varphi,-\frac{\cos\varphi}{\tan\chi \sin\theta} \right)\,, \nn \\
X^\mu_6&=\left( 0,-\cos \theta,\cot \chi \sin \theta,0\right)\,.
\end{align}
The commutators between the generators satisfies the relation
\begin{align}
[J_{ij}, J_{kl}] = \delta_{jk} J_{il} - \delta_{ik} J_{jl} + \delta_{il} J_{jk} - \delta_{jl} J_{ik}\,,
\end{align}
that is nothing but the algebra $\mathfrak{so}(4)$ of the Lie group $SO(4)$.

The $O(4)$ invariance is exploited considering all the possible reflection transformations
\begin{align}\label{refl1}
x_1 \to -x_1 \qquad &\Rightarrow \qquad \varphi \,\to \,\varphi'=\pi -\varphi \quad \, \quad \frac{d r'}{dr}= \frac{d \chi'}{d\chi}=\frac{d \theta'}{d\theta}=1\,,\,\frac{d \varphi'}{d\varphi}=-1\,, \\
\label{refl2}
x_2 \to -x_2 \qquad &\Rightarrow \qquad \phi \,\to\, \phi'=-\phi \quad \, \quad \frac{d r'}{dr}= \frac{d \chi'}{d\chi}=\frac{d \theta'}{d\theta}=1\,,\,\frac{d \varphi'}{d\varphi}=-1\,, \\
\label{refl3}
x_3 \to -x_3 \qquad &\Rightarrow \qquad \theta \,\to\, \theta'=\pi-\theta \quad \, \quad \frac{d r'}{dr}= \frac{d \chi'}{d\chi}=\frac{d \varphi'}{d\varphi}=1\,,\,\frac{d \theta'}{d\theta}=-1\,,  \\
\label{refl4}
x_4 \to -x_4 \qquad &\Rightarrow \qquad \chi \,\to\, \chi'=\pi-\chi \quad \, \quad \frac{d r'}{dr}=\frac{d \theta'}{d\theta}=\frac{d \varphi'}{d\varphi}=1\,,\,\frac{d \chi'}{d\chi}=-1 \,,
\end{align}
from which we read the following rules for the linear affine connection: (i) if the number of coordinate indices $\chi$, $\theta$ or $\varphi$ on $\Gamma^\alpha{}_{\mu\nu}$ is odd, a factor $-1$ is incurred, (ii) additional factors $-1$ arises depending on the parity under such transformations of the trigonometric functions of $\chi$, $\theta$ and $\varphi$.


\begin{thebibliography}{100}

%\bibitem{cab}   N. Cabibbo, L. Maiani, G. Parisi, R. Petronzio, Nucl. Phys. B158 (1979) 295.

%\cite{Krasnikov:1978pu}
\bibitem{Krasnikov:1978pu}
N.~V.~Krasnikov,
%``Restriction of the Fermion Mass in Gauge Theories of Weak and Electromagnetic Interactions,''
Yad. Fiz. \textbf{28} (1978), 549-551.
%100 citations counted in INSPIRE as of 06 Mar 2026

%\cite{Cabibbo:1979ay}
\bibitem{Cabibbo:1979ay}
N.~Cabibbo, L.~Maiani, G.~Parisi and R.~Petronzio,
%``Bounds on the Fermions and Higgs Boson Masses in Grand Unified Theories,''
Nucl. Phys. B \textbf{158} (1979), 295-305.
%doi:10.1016/0550-3213(79)90167-6
%874 citations counted in INSPIRE as of 05 Mar 2026

%\cite{Hung:1979dn}
\bibitem{Hung:1979dn}
P.~Q.~Hung,
%``Vacuum Instability and New Constraints on Fermion Masses,''
Phys. Rev. Lett. \textbf{42} (1979), 873.
%doi:10.1103/PhysRevLett.42.873
%311 citations counted in INSPIRE as of 06 Mar 2026

%\cite{Politzer:1978ic}
\bibitem{Politzer:1978ic}
H.~D.~Politzer and S.~Wolfram,
%``Bounds on Particle Masses in the Weinberg-Salam Model,''
Phys. Lett. B \textbf{82} (1979), 242-246
[erratum: Phys. Lett. B \textbf{83} (1979), 421].
%237 citations counted in INSPIRE as of 06 Mar 2026

%\cite{Anselm:1980mj}
\bibitem{Anselm:1980mj}
A.~A.~Anselm,
%``Problem of Particle Families and Compound SU(5) Decuplets,''
JETP Lett. \textbf{31} (1980), 138
LENINGRAD-80-546.
%7 citations counted in INSPIRE as of 06 Mar 2026

\bibitem{turner} M.S. Turner and F. Wilczek, Nature 298 (1982) 633.
\bibitem{rees} P. Hut and M-J. Rees, Nature 302 (1983) 508.

\bibitem{sher} R.A. Flores, M. Sher, Phys. Rev. D27 (1983) 1679.

%\bibitem{lindn} M. Lindner, Z. Phys. 31 (1986) 295.

%\cite{Lindner:1985uk}
\bibitem{Lindner:1985uk}
M.~Lindner,
%``Implications of Triviality for the Standard Model,''
Z. Phys. C \textbf{31} (1986), 295.
%doi:10.1007/BF01479540
%521 citations counted in INSPIRE as of 05 Mar 2026

\bibitem{sherrep} M. Sher, Phys. Rep. 179 (1989) 273.

%\bibitem{zagla} M. Lindner, M. Sher, H. W. Zaglauer, Phys. Lett. B228 (1989) 139.

%\cite{Lindner:1988ww}
\bibitem{Lindner:1988ww}
M.~Lindner, M.~Sher and H.~W.~Zaglauer,
%``Probing Vacuum Stability Bounds at the Fermilab Collider,''
Phys. Lett. B \textbf{228} (1989), 139-143.
%doi:10.1016/0370-2693(89)90540-6
%299 citations counted in INSPIRE as of 05 Mar 2026

%\bibitem{jones} C. Ford, D.R.T. Jones, P.W. Stephenson, M.B. Einhorn, Nucl. Phys. B395 (1993) 17.

%\cite{Ford:1992mv}
\bibitem{Ford:1992mv}
C.~Ford, D.~R.~T.~Jones, P.~W.~Stephenson and M.~B.~Einhorn,
%``The Effective potential and the renormalization group,''
Nucl. Phys. B \textbf{395} (1993), 17-34.
%doi:10.1016/0550-3213(93)90206-5
%[arXiv:hep-lat/9210033 [hep-lat]].
%424 citations counted in INSPIRE as of 05 Mar 2026

%\bibitem{sher2} M. Sher, Phys. Lett. B317 (1993) 159.

%\cite{Sher:1993mf}
\bibitem{Sher:1993mf}
M.~Sher,
%``Precise vacuum stability bound in the standard model,''
Phys. Lett. B \textbf{317} (1993), 159-163.
%doi:10.1016/0370-2693(93)91586-C
%[arXiv:hep-ph/9307342 [hep-ph]].
%266 citations counted in INSPIRE as of 05 Mar 2026

%\cite{Altarelli:1994rb}
\bibitem{Altarelli:1994rb}
G.~Altarelli and G.~Isidori,
%``Lower limit on the Higgs mass in the standard model: An Update,''
Phys. Lett. B \textbf{337} (1994), 141-144.
%doi:10.1016/0370-2693(94)91458-3
%463 citations counted in INSPIRE as of 05 Mar 2026

%\bibitem{alta}  G. Altarelli, G. Isidori, Phys. Lett. B337 (1994) 141.
\bibitem{quiro} J.A. Casas, J.R. Espinosa, M. Quir\'{o}s, 
Phys. Lett. B342 (1995) 171; Phys. Lett. B382 (1996) 374.
\bibitem{isido} G. Isidori, G. Ridolfi, A. Strumia, 
                Nucl. Phys. B609 (2001) 387.
\bibitem{Espinosa:2007qp}
{}J.~R.~Espinosa, G.~F.~Giudice and A.~Riotto,
JCAP {0805} (2008) 002.
\bibitem{lee} B.~H.~Lee and W.~Lee, Class.Quant.Grav. 
26 (2009) 225002; B.~H.~Lee, W.~Lee, C.~Oh, D.~Ro and 
D.~h.~Yeom, JHEP 1306 (2013) 003; B.~H.~Lee, W.~Lee, 
D.~Ro and D.~h.~Yeom, Phys.Rev. D91 (2015), 124044. 

\bibitem{higgsmass} G. Aad et al. (ATLAS Collaboration, 
CMS Collaboration), Phys. Rev. Lett. 114 (2015) 191803.
\bibitem{ATLAS:2014wva} ATLAS and CDF and CMS and D0 
Collaborations, ``First combination of Tevatron and LHC 
measurements of the top-quark mass'', arXiv:1403.4427 [hep-ex].

\bibitem{isidue} G. Degrassi, S. Di Vita, J. Elias-Miro, 
J.R. Espinosa, 
G.F. Giudice, G. Isidori, A. Strumia, JHEP 1208 (2012) 098.
\bibitem{millington1} 
B. Garbrecht and P. Millington, Phys. Rev. D91 (2015) 105021. 
\bibitem{grinstein} 
B. Grinstein and C.W. Murphy,  JHEP 1512 (2015) 063.
\bibitem{millington2} 
B. Garbrecht and P. Millington, J. Phys. Conf. Ser. 873 (2017) 012041.

\bibitem{raja1} 
M.~Herranen, T.~Markkanen, S.~Nurmi and A.~Rajantie,
Phys.Rev.Lett. {\bf 113} (2014) 211102.
\bibitem{khan} 		
N. Khan and S. Rakshit, Phys. Rev. D90 (2014) 113008.
\bibitem{raja2} 
M.~Herranen, T.~Markkanen, S.~Nurmi and A.~Rajantie,
Phys.Rev.Lett. {\bf 115} (2015) 241301. 
\bibitem{kearney} 	
J. Kearney, H. Yoo  and K. M. Zurek, Phys. Rev. D91 (2015) 123537.
\bibitem{goldberg} 	
L.A. Anchordoqui, V. Barger, H. Goldberg, X. Huang, D. Marfatia, L.H.M. da Silva and T. J. Weiler,
Phys. Rev. D92 (2015) 063504.
\bibitem{macdonald} 
F. Kahlhoefer and John McDonald, JCAP 1511 (2015) 015.
\bibitem{ema1} 
Y. Ema, K. Mukaida and K. Nakayama, JCAP 1610 (2016) 043 
\bibitem{ema2} 
Y. Ema, K. Mukaida and K. Nakayama,  Phys. Lett.  B761 (2016) 419. 
\bibitem{okada} 	
N. Okada and D. Raut, Eur. Phys. J.  C77 (2017) 247.
\bibitem{urbanowski1} 
K. Urbanowski, Theor. Math. Phys. 190 (2017) 458.
\bibitem{urbanowski2} 
A. Stachowski, M. Szydłowski and K. Urbanowski, Eur. Phys. J. C77 (2017) 357.
\bibitem{bu} D. Buttazzo, G. Degrassi, P. P. Giardino, 
G. F. Giudice, F. Sala, A. Salvio, A. Strumia, JHEP
1312 (2013) 089.
\bibitem{NNLO} L.N. Mihaila, J. Salomon and M. Steinhauser, Phys. Rev. Lett. 108 (2012) 151602; K. Chetyrkin and M. Zoller, JHEP 1206 (2012) 033;
F. Bezrukov, M. Yu. Kalmykov, B. A. Kniehl, M. Shaposhnikov, 
JHEP 1210 (2012) 140.
\bibitem{isiuno} J. Elias-Miro, J.R. Espinosa, G.F. Giudice, G. 
Isidori, A. Riotto, A. Strumia, Phys. Lett. B709 (2012) 222. 

\bibitem{Branchina:2013jra}
V.~Branchina and E.~Messina,
%``Stability, Higgs Boson Mass and New Physics,''
Phys. Rev. Lett. \textbf{111} (2013), 241801. %, arXiv:1307.5193.

\bibitem{Branchina:2014usa}
V.~Branchina, E.~Messina and A.~Platania,
%``Top mass determination, Higgs inflation, and vacuum stability,''
JHEP \textbf{09} (2014), 182. %, arXiv:1407.4112.

\bibitem{Branchina:2014rva}
V.~Branchina, E.~Messina and M.~Sher,
%``Lifetime of the electroweak vacuum and sensitivity to Planck scale physics,''
Phys. Rev. D \textbf{91} (2015), 013003. %, arXiv:1408.5302.

\bibitem{haba} 
N. Haba, H. Ishida, R. Takahashi and  Y. Yamaguchi, 
Nucl. Phys. B900 (2015) 244.
\bibitem{ferreira} 
P. M. Ferreira and B. Swiezewska,  JHEP 1604 (2016) 099.

\bibitem{Branchina:2015nda}
V.~Branchina and E.~Messina,
%``Stability and UV completion of the Standard Model,''
EPL \textbf{117} (2017) 6, 61002. %, arXiv:1507.08812.

\bibitem{Branchina:2018qlf}
V.~Branchina, F.~Contino and P.~M.~Ferreira,
%``Electroweak vacuum lifetime in two Higgs doublet models,''
JHEP \textbf{11} (2018), 107. %, arXiv:1807.10802.

\bibitem{Branchina:2016bws}
V.~Branchina, E.~Messina and D.~Zappala,
%``Impact of Gravity on Vacuum Stability,''
EPL \textbf{116} (2016) 2, 21001. %, arXiv:1601.06963.

\bibitem{chaka1} 	
N. Chakrabarty and B. Mukhopadhyaya, Eur. Phys. J. C77 (2017) 153.

\bibitem{Bentivegna:2017qry}
E.~Bentivegna, V.~Branchina, F.~Contino and D.~Zappal{\`a},
%``Impact of New Physics on the EW vacuum stability in a curved spacetime background,''
JHEP \textbf{12} (2017), 100. %, arXiv:1708.01138.

\bibitem{Branchina:2018xdh}
V.~Branchina, F.~Contino and A.~Pilaftsis,
%``Protecting the stability of the electroweak vacuum from Planck-scale gravitational effects,''
Phys. Rev. D \textbf{98} (2018) 7, 075001 %, arXiv:1806.11059.

\bibitem{Branchina:2019tyy}
V.~Branchina, E.~Bentivegna, F.~Contino and D.~Zappal{\`a},
%``Direct Higgs-gravity interaction and stability of our Universe,''
Phys. Rev. D \textbf{99} (2019) 9, 096029. %, arXiv:1905.02975.

\bibitem{bender} T. Banks, C. M. Bender, T. T. Wu, 
Phys. Rev. D8 (1973) 3346; Phys. Rev. D8 (1973) 3366.


%\cite{Lee:1974ma}
\bibitem{Lee:1974ma}
T.~D.~Lee and G.~C.~Wick,
%``Vacuum Stability and Vacuum Excitation in a Spin 0 Field Theory,''
Phys. Rev. D \textbf{9} (1974), 2291-2316.
%832 citations counted in INSPIRE as of 03 Mar 2026

%\cite{Zeldovich:1974uw}
%\bibitem{Zeldovich:1974uw}
%Y.~B.~Zeldovich, I.~Y.~Kobzarev and L.~B.~Okun,
%``Cosmological Consequences of the Spontaneous Breakdown of Discrete Symmetry,''
%Zh. Eksp. Teor. Fiz. \textbf{67} (1974), 3-11 SLAC-TRANS-0165, (Sov.Phys.JETP 40 (1974) 1-5).
%1111 citations counted in INSPIRE as of 03 Mar 2026


%\cite{Kobzarev:1974cp}
\bibitem{Kobzarev:1974cp}
I.~Y.~Kobzarev, L.~B.~Okun and M.~B.~Voloshin,
%``Bubbles in Metastable Vacuum,''
Yad. Fiz. \textbf{20} (1974), 1229-1234
ITEP-81-1974.
%481 citations counted in INSPIRE as of 03 Mar 2026

%\cite{Stone:1975bd}
\bibitem{Stone:1975bd}
M.~Stone,
%``The Lifetime and Decay of Excited Vacuum States of a Field Theory Associated with Nonabsolute Minima of Its Effective Potential,''
Phys. Rev. D \textbf{14} (1976), 3568.
%doi:10.1103/PhysRevD.14.3568
%85 citations counted in INSPIRE as of 03 Mar 2026

%\cite{Stone:1976qh}
\bibitem{Stone:1976qh}
M.~Stone,
%``Semiclassical Methods for Unstable States,''
Phys. Lett. B \textbf{67} (1977), 186-188.
%doi:10.1016/0370-2693(77)90099-5
%70 citations counted in INSPIRE as of 03 Mar 2026

%\cite{Frampton:1976pb}
\bibitem{Frampton:1976pb}
P.~H.~Frampton,
%``Consequences of Vacuum Instability in Quantum Field Theory,''
Phys. Rev. D \textbf{15} (1977), 2922.
%doi:10.1103/PhysRevD.15.2922
%94 citations counted in INSPIRE as of 03 Mar 2026

\bibitem{coleman} S. Coleman, Phys. Rev. D15 (1977) 2929; 
C. G. Callan, S. Coleman, Phys. Rev. D16 (1977) 1762.

%\cite{Linde:1977mm}
\bibitem{Linde:1977mm}
A.~D.~Linde,
%``On the Vacuum Instability and the Higgs Meson Mass,''
Phys. Lett. B \textbf{70} (1977), 306-308.
%doi:10.1016/0370-2693(77)90664-5
%349 citations counted in INSPIRE as of 03 Mar 2026

\bibitem{cdl} S. Coleman, F. De Luccia, Phys. Rev. D21 (1980) 3305.

 
\bibitem{Misner:1973prb}
C.~W.~Misner, K.~S.~Thorne and J.~A.~Wheeler,
%``Gravitation,''
W. H. Freeman, 1973,
ISBN 978-0-7167-0344-0, 978-0-691-17779-3.

\bibitem{Schiff:1960ggq}
L.~I.~Schiff,
%``On Experimental Tests of the General Theory of Relativity,''
Am. J. Phys. \textbf{28} (1960) no.4, 340.
%doi:10.1119/1.1935800

\bibitem{Coley82}
A.~Coley
Phys. Rev. Lett. \textbf{49} (1982)
American Physical Society.


\bibitem{Capozziello:2025hyw}
S.~Capozziello, S.~Cesare and C.~Ferrara,
Eur. Phys. J. C \textbf{85} (2025) no.9, 932.


\bibitem{Bonola1955}
R.~Bonola, Dover Books on Mathematics Series, Dover Publications, 1955.

\bibitem{Bennett2011}
M.~K.~Bennett,
Wiley, 2011.


%\cite{Sotiriou:2008rp}
\bibitem{Sotiriou:2008rp}
T.~P.~Sotiriou and V.~Faraoni,
Rev. Mod. Phys. \textbf{82} (2010), 451-497


\bibitem{Hooft1974}
G.~'t~Hooft and M.~Veltman,
\textit{Ann. Inst. H. Poincar\'e Phys. Th\'eor.}
\textbf{20}, no.~1, 69-94 (1974).

\bibitem{Obukhov_2017}
Y.~N.~Obukhov,
\textit{Phys. Rev. D} \textbf{95}, no.~8, 084028 (2017).


\bibitem{Giulini2003}
D.~Giulini, C.~Kiefer, and C.~L\"ammerzahl,
vol.~631 (Springer, 2003).


\bibitem{Rovelli2004}
C.~Rovelli,
Cambridge Monographs on Mathematical Physics
(Cambridge University Press, Cambridge, 2004).


\bibitem{Perivolaropoulos:2021jda}
L.~Perivolaropoulos and F.~Skara,
\textit{New Astron. Rev.} \textbf{95}, 101659 (2022).


\bibitem{DESI:2024mwx}
A.~G.~Adame et al. (DESI Collaboration),
\textit{JCAP} \textbf{02}, 021 (2025).


\bibitem{Mayet:2016zxu}
F.~Mayet et al.,
\textit{Phys. Rept.} \textbf{627}, 1-49 (2016).


\bibitem{Capozziello:2017rvz}
S.~Capozziello, P.~Jovanovi\'c, V.~B.~Jovanovi\'c, and D.~Borka,
\textit{JCAP} \textbf{06}, 044 (2017).


\bibitem{Arun_2017}
K.~Arun, S.~B.~Gudennavar, and C.~Sivaram,
\textit{Adv. Space Res.} \textbf{60}, no.~1, 166-186 (2017).

\bibitem{Carroll:2000fy}
S.~M.~Carroll,
\textit{Living Rev. Rel.} \textbf{4}, 1 (2001).

\bibitem{Planck:2018vyg}
N.~Aghanim et al. (Planck Collaboration),
\textit{Astron. Astrophys.} \textbf{641}, A6 (2020).


\bibitem{Guth:2005zr}
A.~H.~Guth and D.~I.~Kaiser,
\textit{Science} \textbf{307}, 884-890 (2005).


\bibitem{Pantos:2026koc}
I.~Pantos and L.~Perivolaropoulos,
\textit{Phys. Dark Univ.} \textbf{52}, 102286 (2026).


\bibitem{Aloni:2021eaq}
D.~Aloni, A.~Berlin, M.~Joseph, M.~Schmaltz, and N.~Weiner,
\textit{Phys. Rev. D} \textbf{105}, no.~12, 123516 (2022).

\bibitem{DiValentino:2021izs}
E.~Di~Valentino, O.~Mena, S.~Pan, L.~Visinelli, W.~Yang,
A.~Melchiorri, D.~F.~Mota, A.~G.~Riess, and J.~Silk,
\textit{Class. Quant. Grav.} \textbf{38}, no.~15, 153001 (2021).


\bibitem{Coley:2019zld}
A.~A.~Coley, R.~J.~Van Den Hoogen and D.~D.~McNutt,
J. Math. Phys. \textbf{61} (2020) no.7, 072503.


\bibitem{Bahamonde:2015zma}
S.~Bahamonde, C.~G.~B{\"o}hmer and M.~Wright,
Phys. Rev. D \textbf{92} (2015) no.10, 104042.

\bibitem{BeltranJimenez:2019esp}
J.~Beltr{\'a}n Jim{\'e}nez, L.~Heisenberg and T.~S.~Koivisto,
Universe \textbf{5} (2019) 7, 173.

\bibitem{Capozziello:2022zzh}
S.~Capozziello, V.~De Falco and C.~Ferrara,
Eur. Phys. J. C \textbf{82} (2022) 10, 865.


%\cite{Battista:2026gvo}
\bibitem{Battista:2026gvo}
E.~Battista, S.~Capozziello and S.~Pastore,
Eur. Phys. J. C \textbf{86}, no.3, 298 (2026).

\bibitem{Capozziello:2026pys}
S.~Capozziello and G.~Meluccio,
%``The pre-geometric origin of geometric trinity of gravity,''
Eur. Phys. J. C \textbf{86}  no.9, 1073 (2026).

\bibitem{Pereira:2013qza}
J.~G.~Pereira,
arXiv:1302.6983 [gr-qc].
%\cite{Iosifidis:2023eom}
\bibitem{Iosifidis:2023eom}
D.~Iosifidis and F.~W.~Hehl,
Phys. Lett. B \textbf{850} (2024), 138498.


\bibitem{Iosifidis(2026)}
D. Iosifidis, arXiv:2604.22845 [physics.gen-ph]. 


%\cite{Brezina:2025dbc}
\bibitem{Brezina:2025dbc}
S.~Brezina, E.~Boffo and M.~Kr{\v{s}}{\v{s}}{\'a}k,
JHEP \textbf{05} (2026), 182.

\bibitem{Tino:2020nla}
G.~M.~Tino, L.~Cacciapuoti, S.~Capozziello, G.~Lambiase and F.~Sorrentino,
Prog. Part. Nucl. Phys. \textbf{112} (2020), 103772.



\bibitem{Alonso:2022oot}
I.~Alonso, C.~Alpigiani, B.~Altschul, H.~Ara{\'u}jo, G.~Arduini, J.~Arlt, L.~Badurina, A.~Bala{\v{z}}, S.~Bandarupally and B.~C.~Barish, \textit{et al.}
EPJ Quant. Technol. \textbf{9} (2022) no.1, 30.

\bibitem{Altschul:2014lua}
B.~Altschul, Q.~G.~Bailey, L.~Blanchet, K.~Bongs, P.~Bouyer, L.~Cacciapuoti, S.~Capozziello, N.~Gaaloul, D.~Giulini and J.~Hartwig, \textit{et al.}
Adv. Space Res. \textbf{55} (2015), 501-524.

\bibitem{Capozziello:2024ijv}
S.~Capozziello and C.~Ferrara,
Int. J. Geom. Meth. Mod. Phys. \textbf{21} (2024) no.10, 2440014.

\bibitem{Will1993}
C.~M.~Will,
Cambridge University Press, 2018,
ISBN 978-1-108-67982-4, 978-1-107-11744-0.

\bibitem{Faraoni2010}
V.~Faraoni and S.~Capozziello,
Springer, 2011,
ISBN 978-94-007-0164-9, 978-94-007-0165-6.


\bibitem{Bahamonde:2021gfp}
S.~Bahamonde, K.~F.~Dialektopoulos, C.~Escamilla-Rivera, G.~Farrugia, V.~Gakis, M.~Hendry, M.~Hohmann, J.~Levi Said, J.~Mifsud and E.~Di Valentino,
Rept. Prog. Phys. \textbf{86} (2023) 2, 026901.

\bibitem{BeltranJimenez:2017tkd}
J.~Beltr{\'a}n Jim{\'e}nez, L.~Heisenberg and T.~Koivisto,
Phys. Rev. D \textbf{98} (2018) 4, 044048.

\bibitem{Mancini:2025asp}
C.~Mancini, G.~M.~Tino and S.~Capozziello,
Found. Phys. \textbf{55} (2025) no.5, 69.


\bibitem{Hohmann:2019nat}
M.~Hohmann, L.~J{\"a}rv, M.~Kr{\v{s}}{\v{s}}{\'a}k and C.~Pfeifer,
Phys. Rev. D \textbf{100} (2019) 8, 084002.

\bibitem{Blixt:2018znp}
D.~Blixt, M.~Hohmann and C.~Pfeifer,
Phys. Rev. D \textbf{99} (2019) no.8, 084025.


\bibitem{Capozziello:2021pcg}
S.~Capozziello, A.~Finch, J.~L.~Said and A.~Magro,
Eur. Phys. J. C \textbf{81} (2021) no.12, 1141.


\bibitem{DAmbrosio:2021zpm}
F.~D'Ambrosio, S.~D.~B.~Fell, L.~Heisenberg and S.~Kuhn,
Phys. Rev. D \textbf{105} (2022) no.2, 024042.



\bibitem{Jarv:2023sbp}
L.~Jarv and L.~Pati,
Phys. Rev. D \textbf{109} (2024) no.6, 064069.


\bibitem{Golovnev:2017dox}
A.~Golovnev, T.~Koivisto and M.~Sandstad,
Class. Quant. Grav. \textbf{34} (2017) 14, 145013.

\bibitem{Pereira:2012kd}
J.~G.~Pereira,
AIP Conf. Proc. \textbf{1483} (2012) no.1, 239-259.


\bibitem{Krssak:2015rqa}
M.~Kr{\v{s}}{\v{s}}{\'a}k and J.~G.~Pereira,
Eur. Phys. J. C \textbf{75} (2015) no.11, 519.

\bibitem{Krssak:2018ywd}
M.~Krssak, R.~J.~van den Hoogen, J.~G.~Pereira, C.~G.~B{\"o}hmer and A.~A.~Coley,
Class. Quant. Grav. \textbf{36} (2019) no.18, 183001.


\bibitem{Hehl:1994ue}
F.~W.~Hehl, J.~D.~McCrea, E.~W.~Mielke and Y.~Ne'eman,
Phys. Rept. \textbf{258} (1995), 1-171.

\bibitem{Hayashi:1979qx}
K.~Hayashi and T.~Shirafuji,
Phys. Rev. D \textbf{19} (1979), 3524-3553.


\bibitem{Obukhov:2004hv}
Y.~N.~Obukhov and J.~G.~Pereira,
Phys. Rev. D \textbf{69} (2004), 128502.


\bibitem{Aldrovandi:2013wha}
R.~Aldrovandi and J.~G.~Pereira,
Springer, 2013,
ISBN 978-94-007-5142-2, 978-94-007-5143-9.

\bibitem{BeltranJimenez:2018vdo}
J.~Beltr{\'a}n Jim{\'e}nez, L.~Heisenberg and T.~S.~Koivisto,
JCAP \textbf{08} (2018), 039.

\bibitem{Capozziello:2023vne}
S.~Capozziello, V.~De Falco and C.~Ferrara,
Eur. Phys. J. C \textbf{83} (2023) no.10, 915.


\bibitem{Hohmann:2022mlc}
M.~Hohmann,
Lect. Notes Phys. \textbf{1017} (2023), 145.

\bibitem{Heisenberg:2023lru}
L.~Heisenberg,
Phys. Rept. \textbf{1066} (2024), 1-78.

\bibitem{Oshita:2017nhn}
N.~Oshita and Y.~P.~Wu,
Phys. Rev. D \textbf{96} (2017) 4, 044042.%, arXiv:1705.10436.

\bibitem{BeltranJimenez:2020sih}
J.~Beltr{\'a}n Jim{\'e}nez, L.~Heisenberg and T.~Koivisto,
Class. Quant. Grav. \textbf{37} (2020) no.19, 195013.
%doi:10.1088/1361-6382/aba31b[arXiv:2004.04606 [hep-th]].

\bibitem{Hohmann:2024phz}
M.~Hohmann and V.~Karanasou,
Phys. Rev. D \textbf{111} (2025) no.6, 064057.
%doi:10.1103/PhysRevD.111.064057 [arXiv:2412.11730 [gr-qc]].


\bibitem{Hohmann:2019fvf}
M.~Hohmann,
Symmetry \textbf{12} (2020) no.3, 453.

\bibitem{Ferrara:2026hbq}
C.~Ferrara, A.~Golovnev and M.~J.~Guzm{\'a}n,
Int. J. Geom. Meth. Mod. Phys. \textbf{23} (2026), 2650297
doi:10.1142/S021988782650297X
[arXiv:2605.30221 [gr-qc]].


\bibitem{Blixt:2023kyr}
D.~Blixt, A.~Golovnev, M.~J.~Guzman and R.~Maksyutov,
Phys. Rev. D \textbf{109} (2024) no.4, 044061

\bibitem{Ferrara:2026bzy}
C.~Ferrara, M.~J.~Guzm{\'a}n and L.~J{\"a}rv,
[arXiv:2609.39988 [gr-qc]].




\end{thebibliography}
\end{document}